\documentclass[final]{siamltex}

\usepackage{amssymb,latexsym,amsmath}
\usepackage{epsfig}
\usepackage{caption}
\usepackage{caption}
 
 \usepackage{epsfig}

\usepackage{makeidx}

\allowdisplaybreaks

 \usepackage{graphicx,fancybox,latexsym,epsfig}
\usepackage{fancyhdr,amsmath,times,amsxtra,amssymb}
\usepackage{color}

\def\Pr{\mathop{\rm Pr}}

\def\P{{\mathcal P}}

\def\sPr{{\mathsf{Pr}}}

\def\sX{{\mathds X}}

\def\sU{{\mathds U}}

\def\sZ{{\mathds Z}}

\def\sU{{\mathds U}}

\usepackage{amsmath}
\usepackage{amsfonts}
\usepackage{array}
\usepackage{dsfont}
\usepackage{hyperref}
\usepackage{amssymb}
\usepackage{bbold}
\usepackage{caption}
\usepackage{standalone}
\usepackage{accents}
\usepackage{enumitem}
\usepackage{tikz}
\usepackage{pgfplots}
\pgfplotsset{compat=1.6}
\usepgfplotslibrary{groupplots}
\usepackage{cancel}
\allowdisplaybreaks

\newtheorem{assumption}{Assumption}[section]
\newtheorem{exmp}{Example}[section]

\newtheorem{mydef}{Definition}
  \newtheorem{remark}{Remark}[section]
  \newtheorem{example}{Example}[section]

\usepackage{amssymb,latexsym,amsmath}
\usepackage{mathrsfs}
\usepackage{epsfig}
\usepackage{caption}

\newcommand{\R}{\mathds{R}}
\newcommand{\Zplus}{\mathds{Z}_+}
\newcommand{\N}{\mathds{N}}

\newcommand{\dd}{\mathrm{d}}

\pgfplotsset{soldot/.style={color=blue,only marks,mark=*}}
\pgfplotsset{holdot/.style={color=blue,fill=white,only marks,mark=*}}
\fussy

\allowdisplaybreaks

\begin{document}
\sloppy
\title{Robustness to Incorrect System Models in Stochastic Control\thanks{
This research was supported in part by
the Natural Sciences and Engineering Research Council (NSERC) of Canada. Some of the results in this paper were presented at the 2018 IEEE Conference on Decision and Control. The paper is to appear in SIAM J. on Control and Optimization.}
}
\author{Al\.{I} Devran Kara and Serdar Y\"uksel
\thanks{The authors are with the Department of Mathematics and Statistics,
     Queen's University, Kingston, ON, Canada,
     Email: \{16adk,yuksel\}@queensu.ca}
     }
\maketitle
\begin{abstract}
In stochastic control applications, typically only an ideal model (controlled transition kernel) is assumed and the control design is based on the given model, raising the problem of performance loss due to the mismatch between the assumed model and the actual model. Toward this end, we study continuity properties of discrete-time stochastic control problems with respect to system models (i.e., controlled transition kernels) and robustness of optimal control policies designed for incorrect models applied to the true system. We study both fully observed and partially observed setups under an infinite horizon discounted expected cost criterion. We show that continuity can be established under total variation convergence of the transition kernels under mild assumptions and with further restrictions on the dynamics and observation model under weak and setwise convergence of the transition kernels. Using these continuity properties, we establish convergence results and error bounds due to mismatch that occurs by the application of a control policy which is designed for an incorrectly estimated system model to a true model, thus establishing positive and negative results on robustness. Compared to the existing literature, we obtain strictly refined robustness results that are applicable even when the incorrect models can be investigated under weak convergence and setwise convergence criteria (with respect to a true model), in addition to the total variation criteria. These entail positive implications on empirical learning in (data-driven) stochastic control since often system models are learned through empirical training data where typically weak convergence criterion applies but stronger convergence criteria do not. 
 \end{abstract}


\section{Introduction}\label{section:intro}
\subsection{Preliminaries}
In this paper, we study continuity properties of stochastic control problems with respect to transition kernels and applications of these to robustness of optimal control policies applied to systems with incomplete or incorrect probabilistic models.


We start with the probabilistic setup of the problem. Let $\mathds{X} \subset \mathds{R}^m$ denote a Borel set which is the state space of a partially observed controlled Markov process. Here and throughout the paper $\Zplus$ denotes the set of non-negative
integers and $\mathds{N}$ denotes the set of positive integers. Let
$\mathds{Y} \subset \mathds{R}^n$ be a Borel set denoting the observation space of the model, and let the state be observed through an
observation channel $Q$. The observation channel, $Q$, is defined as a stochastic kernel (regular
conditional probability) from  $\mathds{X}$ to $\mathds{Y}$, such that
$Q(\,\cdot\,|x)$ is a probability measure on the (Borel)
$\sigma$-algebra ${\cal B}(\mathds{Y})$ of $\mathds{Y}$ for every $x
\in \mathds{X}$, and $Q(A|\,\cdot\,): \mathds{X}\to [0,1]$ is a Borel
measurable function for every $A \in {\cal B}(\mathds{Y})$.  A
decision maker (DM) is located at the output of the channel $Q$, hence it only sees the observations $\{Y_t,\, t\in \Zplus\}$ and chooses its actions from $\mathds{U}$, the action space which is a Borel
subset of some Euclidean space. An {\em admissible policy} $\gamma$ is a
sequence of control functions $\{\gamma_t,\, t\in \Zplus\}$ such
that $\gamma_t$ is measurable with respect to the $\sigma$-algebra
generated by the information variables
\[
I_t=\{Y_{[0,t]},U_{[0,t-1]}\}, \quad t \in \mathds{N}, \quad
  \quad I_0=\{Y_0\},
\]
where
\begin{equation}
\label{eq_control}
U_t=\gamma_t(I_t),\quad t\in \Zplus
\end{equation}
are the $\mathds{U}$-valued control
actions and 
\[Y_{[0,t]} = \{Y_s,\, 0 \leq s \leq t \}, \quad U_{[0,t-1]} =
  \{U_s, \, 0 \leq s \leq t-1 \}.\]

\noindent We define $\Gamma$ to be the set of all such admissible policies.

The update rules of the system are determined by (\ref{eq_control}) and the following
relationships:
\[  \Pr\bigl( (X_0,Y_0)\in B \bigr) =  \int_B P(dx_0)Q(dy_0|x_0), \quad B\in \mathcal{B}(\mathds{X}\times\mathds{Y}), \]
where $P$ is the (prior) distribution of the initial state $X_0$, and
\begin{eqnarray*}
\label{eq_evol}
 \Pr\biggl( (X_t,Y_t)\in B \, \bigg|\, (X,Y,U)_{[0,t-1]}=(x,y,u)_{[0,t-1]} \biggr)
 \\
 = \int_B \mathcal{T}(dx_t|x_{t-1}, u_{t-1})Q(dy_t|x_t),  B\in \mathcal{B}(\mathds{X}\times
\mathds{Y}), t\in \mathds{N},
\end{eqnarray*}
where $\mathcal{T}$ is the transition kernel of the model which is a stochastic kernel from $\mathds{X}\times
\mathds{U}$ to $\mathds{X}$.

Using stochastic realization results (see Lemma 1.2 in \cite{gihman2012controlled}, or Lemma 3.1 of \cite{BorkarRealization}), the process defined above can be represented in functional form as follows:
\begin{eqnarray}\label{controlledMarkovS}
X_{t+1} = f(X_t,U_t, W_t), \qquad  \qquad Y_t= g(X_t,V_t)
\end{eqnarray}
for some measurable functions $f, g$, with
$\{W_t\}$ being an independent and identically distributed (i.i.d) system
noise process and $\{V_t\}$ an i.i.d. disturbance process, which are
independent of $X_0$ and each other. Here, the first equation represents the transition kernel $\mathcal{T}$ as it gives the relation of the most recent state and action variables to the upcoming state. From this representation it can be seen that the probabilistic nature of the kernel is determined by the function $f$ and the probability model of the noise $W_t$. The second equation
represents the measurement channel $Q$, as it describes the relation
between the state and observation variables. We let the objective of the agent (decision maker) be the minimization of the infinite horizon discounted cost, 
  \begin{align*}
    J_{\beta}(P,{\cal T},\gamma)= E_P^{{\cal T},\gamma}\left[\sum_{t=0}^{\infty} \beta^t c(X_t,U_t)\right]
  \end{align*}
 \noindent for some discount factor $\beta \in (0,1)$, over the set of admissible policies $\gamma\in\Gamma$, where $c:\mathds{X}\times\mathds{U}\to\R$ is a Borel-measurable stage-wise cost function and $E_P^{{\cal T},\gamma}$ denotes the expectation with initial state probability measure $P$ and transition kernel ${\cal T}$ under policy $\gamma$. Note that $P\in\mathcal{P}(\mathds{X})$, where we let $\mathcal{P}(\mathds{X})$ denote the set of probability measures on $\mathds{X}$.

We define the optimal cost for the discounted infinite horizon setup as a function of the priors and the transition kernels as
\begin{align*}
  J_{\beta}^*(P,{\cal T})&=\inf_{\gamma\in\Gamma} J_{\beta}(P,{\cal T},\gamma).
\end{align*}
The focus of the paper will be to address the following problems:


\noindent{\bf Problem P1: Continuity of $J_\beta^*(P,\cal{T})$ under the convergence of the transition kernels.}
Let $\{{\cal{T}}_{n},n\in\mathds{N}\}$ be a sequence of transition kernels which converge in some sense to another transition kernel $\cal{T}$. Does that imply that
\begin{align*}
  J_\beta^*(P,{\cal{T}}_n)\to J_\beta^*(P,\cal{T})?
\end{align*}

\noindent{\bf Problem P2: Robustness to incorrect models.} \
 A problem of major practical importance is robustness of an optimal controller to modeling errors. Suppose that an optimal policy is constructed according to a model which is incorrect: how does the application of the control to the true model affect the system performance and does the error decrease to zero as the models become closer to each other? In particular, suppose that $\gamma_n^*$ is an optimal policy designed for ${\cal T}_n$, an incorrect model for a true model ${\cal T}$. Is it the case that if ${\cal T}_n \to {\cal T}$ then $J_{\beta}(P, {\cal T}, \gamma_n^*) \to  J_{\beta}^*(P,{\cal T})$?

\noindent{\bf Problem P3:  Empirical consistency of learned probabilistic models and data-driven stochastic control.}
Let $\mathcal{T}(\cdot|x,u)$ be a transition kernel given previous state and action variables $x\in \mathds{X},u\in\mathds{U}$, which is unknown to the decision maker (DM). Suppose, the DM builds a model for the transition kernels, $\mathcal{T}_n(\cdot|x,u)$, for all possible $x\in \mathds{X}, u \in \mathds{U}$ by collecting training data (e.g. from the evolving system). Do we have that the cost calculated under $\mathcal{T}_n$ converges to the true cost (i.e., do we have that the cost obtained from applying the optimal policy for the empirical model converges to the true cost as the training length increases)?

\subsection{Literature Review}\label{ch:Background}

Robustness is a desired property for the optimal control of stochastic or deterministic systems when a given model does not reflect the actual system perfectly, as is usually the case in practice. 

A common approach in the literature has been to design controllers that works sufficiently well for all possible uncertain systems under some structured constraints, such as $H_\infty$ norm bounded perturbations (see \cite{basbern,zhou1996robust}). The design for robust controllers has often been developed through a game theoretic formulation where the minimizer is the controller and the maximizer is the uncertainty. The connections of this formulation to risk sensitive control were established in \cite{jacobson1973optimal,dupuis2000robust}. Using Legendre-type transforms, relative entropy constraints came in to the literature to probabilistically model the uncertainties, see e.g. \cite[Eqn. (4)]{dai1996connections} or \cite[Eqns. (2)-(3)]{dupuis2000robust}. Here, one selects a nominal system which satisfies a relative entropy bound between the actual measure and the nominal measure, solves a risk sensitive optimal control problem, and this solution value provides an upper bound for the original system performance. As such, a common approach in robust stochastic control has been to consider all models which satisfy certain bounds in terms of relative entropy pseudo-distance (or Kullback-Leibler divergence), see e.g. \cite{dupuis2000robust,dai1996connections,dupuis2000kernel,boel2002robustness} among others. 

Other metrics or criteria, different from the relative entropy pseudo-distance, have also been used to quantify the uncertainty in the system models. 
Reference \cite{tzortzis2015dynamic} has studied a min-max formulation for robust control where the one-stage transition kernel belongs to a ball under the total variation metric for each state action pair. 
For distributionally robust stochastic optimization problems, it is assumed that the underlying probability measure of the system lies within an ambiguity set and a worst case single-stage optimization is made considering the probability measures in the ambiguity set. 
To construct ambiguity sets, \cite{blanchet2016,esfahani2015} use the Wasserstein metric (see Section \ref{def_exmp}), \cite{erdogan2005} uses the Prokhorov metric which metrizes the weak topology, \cite{sun2015} uses the total variation distance and \cite{lam2016} works with relative entropy. \cite{Rustem2012, Iyengar2005, Ghaoui2005} have studied robust dynamic programming approaches through a min-max formulation for fully observed finite state-action space models with uncertain transition probabilities. 
 Further related work with model uncertainty includes \cite{oksendal2014forward, benavoli2011robust,xu_mannor}, with some further work in the economics literature \cite{hansen2001robust, gossner2008entropy}. In \cite{Kleptsyna2016}, an optimal filtering problem for a control-free system with uncertainties in transition kernels and measurement channels is considered.

Further related studies include \cite{Dean18} which studies the optimal control of systems with unknown dynamics for a Linear Quadratic Regulator setup and proposes an algorithm to learn the system from observed data with quantitative convergence bounds. \cite{ugrinovskii1998robust} considers stochastic uncertainties while \cite{savkin1996robust} considers deterministic structured uncertainties in robust control; some connections of these with our paper can be seen in the examples presented in Section \ref{ornek1}. 

For fully observed models, \cite[Theorem 5.1]{Langen81} establishes continuity results for approximate models and gives a set convergence result for sets of optimal control actions, however this set convergence result is inconclusive for robustness without further assumptions on the true system model. Reference \cite{muller1997does} is another related work which studies continuity of the value function for fully observed models under a general metric defined as the integral probability metric which captures both the total variation metric or the Kantorovich metric with different setups (which is not weaker than the metrics leading to weak convergence). We also note that approximation methods for stochastic control problems with standard Borel spaces through quantization, which leads to finite models, can be viewed as approximations of transition kernels, but this interpretation requires caution: indeed, \cite{SaYuLi17, arruda2012, arruda2013}, among many others, study approximation methods for MDP's where the convergence of approximate models is satisfied in a particularly constructed fashion. In particular, \cite{SaYuLi17} presents a construction for the approximate models through quantizing the actual model with continuous spaces (leading to a finite space model), which allows for continuity and robustness results with only a weak continuity assumption on the true transition kernel which, in turn,  leads to the weak convergence of the approximate models. A detailed analysis of approximation methods for continuous state and action spaces can be found in \cite{SLYbook} for both fully observed and partially observed models. However, these positive results on weak convergence of approximate kernels to the true one in such studies do not directly apply to robustness to an arbitrary sequence of models which converges weakly to a true model, as our counterexamples in this paper will demonstrate.

Related work also includes our recent studies \cite{yuksel12:siam,Prior2017}. \cite{yuksel12:siam} considers various topologies on the sets of observation channels and quantizers in partially observed stochastic control and provides some supporting results. \cite{Prior2017} presents robustness and continuity properties for stochastic control problems with respect to the prior measures with fixed transition kernels; that is, the robustness only to initial priors is studied in \cite{Prior2017}. Different from these studies, here we study continuity in and robustness to incorrect transition kernels which requires significantly different analytical tools due to the dynamic nature of the problems.

In Section \ref{contributions} we will present the contributions, which will also make the comparison with the reviewed literature more explicit.

\subsection{Some Examples and Convergence Criteria for Transition Kernels}\label{def_exmp}

\subsubsection{Convergence criteria for transition kernels}
Before presenting convergence criteria for controlled transition kernels, we first review convergence of probability measures. Three important notions of convergences for sets of probability measures to be studied in the paper are weak convergence, setwise convergence and convergence under total variation. For $N\in\N$, a sequence $\{\mu_n,n\in\N\}$ in $\mathcal{P}(\R^N)$ is said to converge to $\mu\in\mathcal{P}(\R^N)$ \emph{weakly} if
  \begin{align*}\label{converge}\tag{$\ast$}
    \int_{\R^N}c(x)\mu_n(\dd x) \to \int_{\R^N}c(x)\mu(\dd x)
  \end{align*}
  \noindent for every continuous and bounded $c:\R^N \to \R$. $\{\mu_n\}$ is said to converge \emph{setwise} to $\mu\in\mathcal{P}(\R^N)$ if (\ref{converge}) holds for all measurable and bounded $c:\R^N \to \R$. 
  For probability measures $\mu,\nu\in\mathcal{P}(\R^N)$, the \emph{total variation} metric is given by
  \begin{align*}
    \|\mu-\nu\|_{TV}=2\sup_{B\in\mathcal{B}(\R^N)}|\mu(B)-\nu(B)| =\sup_{f:\|f\|_\infty \leq 1}|\int f(x)\mu(\dd x)-\int f(x)\nu(\dd x)|,
  \end{align*}
  \noindent where the supremum is taken over all measurable real $f$ such that  $\|f\|_\infty=\sup_{x\in\R^N}|f(x)|\leq 1$. A sequence $\{\mu_n\}$ is said to converge in total variation to $\mu\in\mathcal{P}(\R^N)$ if $\|\mu_n-\mu\|_{TV}\to 0$.
Total variation defines a stringent metric for convergence; for example, a sequence of discrete probability measures does not converge in total variation to a probability measure which admits a density function. Setwise convergence, though, induces a topology on the space of probability measures which is not metrizable \cite[p.~59]{Ghosh}. However, the space of probability measures on a complete, separable, metric (Polish) space endowed with the topology of weak convergence is itself complete, separable and metric \cite{Par67}. We also note here that relative entropy convergence, through Pinsker's inequality \cite[Lemma 5.2.8]{GrayInfo}, is stronger than even total variation convergence which has also been studied in robust stochastic control as reviewed earlier. Another metric for probability measures is the Wasserstein distance: For compact spaces, the Wasserstein distance of order $1$, denoted by $W_1$, metrizes the weak topology (see \cite[Theorem 6.9]{villani2008optimal}). For non-compact spaces convergence in the $W_1$ metric implies weak convergence (in particular this metric bounds from above the Bounded-Lipschitz metric \cite[p.109]{villani2008optimal}). Considering these relations, our results in this paper can be directly generalized to the relative entropy distance or the Wasserstein distance. Building on the above, we introduce the following convergence notions for (controlled) transition kernels.
\begin{mydef}\label{def_kernel}
For a sequence of transition kernels $\{ \mathcal{T}_n,n\in\mathds{N}\}$, we say that
\begin{enumerate}[label=(\roman*)]
\item $\mathcal{T}_n \to  \mathcal{T}$ weakly if $ \mathcal{T}_n(\cdot|x,u) \to  \mathcal{T}(\cdot|x,u)$ weakly, for all $x \in \mathds{X}$ and $u \in \mathds{U}$.
\item $\mathcal{T}_n \to  \mathcal{T}$ setwise if $ \mathcal{T}_n(\cdot|x,u) \to  \mathcal{T}(\cdot|x,u)$ setwise, for all $x \in \mathds{X}$ and $u \in \mathds{U}$.
\item $\mathcal{T}_n \to \mathcal{T}$ under the total variation distance if $ \mathcal{T}_n(\cdot|x,u) \to  \mathcal{T}(\cdot|x,u)$ under total variation, for all $x \in \mathds{X}$ and $u \in \mathds{U}$.
\end{enumerate}
\end{mydef}


\subsubsection{Examples}\label{ornek1}
Let a controlled model be given as
\[x_{t+1}=F(x_t,u_t,w_t),\] where $\{w_t\}$ is an i.i.d. noise process. The uncertainty on the transition kernel for such a system may arise from lack of information on $F$ or the i.i.d. noise process $w_t$ or both:
\begin{itemize}
\item[(i)] Let $\{F_n\}$ denote an approximating sequence for $F$, so that $F_n(x,u,w) \to F(x,u,w)$ pointwise. Assume that the probability measure of the noise is known. Then, corresponding kernels $\mathcal{T}_n$ converges weakly to $\mathcal{T}$: If we denote the probability measure of $w$ with $\mu$, for any $g \in C_b(\mathds{X})$ and for any $(x_0,u_0)\in \mathds{X}\times\mathds{U}$ using the dominated convergence theorem we have
\begin{align*}
&\lim_{n \to \infty}\int g(x_1)\mathcal{T}_n(dx_1|x_0,u_0) =\lim_{n \to \infty}\int g(F_n(x_0,u_0,w)) \mu(dw)\\
&=\int g(F(x_0,u_0,w)) \mu(dw) = \int g(x_1)\mathcal{T}(dx_1|x_0,u_0).
\end{align*}

\item[(ii)] Much of the robust control literature deals with deterministic systems where the nominal model is a deterministic perturbation of the actual model (see e.g. \cite{savkin1996robust,Dean18}). The considered model is in the following form; $\tilde{F}(x_t,u_t) = F(x_t,u_t) + \Delta F(x_t,u_t)$, where $F$ represents the nominal model and $\Delta F$ is the model uncertainty satisfying some norm bounds. For such deterministic systems, pointwise convergence of $\tilde{F}$ to the nominal model $F$, i.e. $\Delta F(x_t,u_t)\to 0$, can be viewed as weak convergence for deterministic systems by the discussion in (i). It is evident, however, that total variation convergence would be too strong for such a convergence criterion, since  $\delta_{\tilde{F}(x_t,u_t)} \to \delta_{F(x_t,u_t)}$ weakly but $\|\delta_{\tilde{F}(x_t,u_t)}-\delta_{F(x_t,u_t)}\|_{TV}=2$ for all $\Delta F(x_t,u_t)\neq 0$.

\item[(iii)]
Let $F(x_t,u_t,w_t) = f(x_t,u_t) + w_t$ be such that the function $f$ is known and $w_t \sim \mu$ is not known correctly and an incorrect model $\mu_n$ is assumed. If $\mu_n \to \mu$ weakly, setwise, or total variation, then the corresponding transition kernels $\mathcal{T}_n$ converges in the same sense to $\mathcal{T}$. Observe the following,
\begin{align}\label{exmp1}
&\int g(x_1)\mathcal{T}_n(dx_1|x_0,u_0) - \int g(x_1)\mathcal{T}(dx_1|x_0,u_0)\nonumber\\
&=\int g(w_0+f(x_0,u_0))\mu_n(dw_0) - \int g(w_0+f(x_0,u_0))\mu(dw_0).
\end{align}
a) Suppose $\mu_n \to \mu$ weakly. If $g$ is a continuous and bounded function then $g(\cdot + f(x_0,u_0))$ is a continuous and bounded function for all $(x_0,u_0)\in \mathds{X}\times\mathds{U}$. Thus, (\ref{exmp1}) goes to 0. Note that $f$ does not need to be continuous. b) Suppose $\mu_n \to \mu$ setwise. If $g$ is a measurable and bounded function, then $g(\cdot +f(x_0,u_0))$ measurable and bounded for all $(x_0,u_0)\in \mathds{X}\times\mathds{U}$. Thus, (\ref{exmp1}) goes to 0. c) Finally, assume  $\mu_n \to \mu$ in total variation. If $g$ is bounded, (\ref{exmp1}) converges to 0, as in item (b). As a special case, assume that $\mu_n$ and $\mu$ admit densities $h_n$ and $h$ respectively; then the pointwise convergence of $h_n$ to $h$ implies the convergence of $\mu_n$ to $\mu$ in total variation by Scheff\'e's Theorem.  
%
%
\item[(iv)]Suppose now neither $F$ nor the probability model of $w_t$ is known perfectly. It is assumed that $w_t$ admits a measure $\mu_n$ and $\mu_n\to \mu$ weakly. For the function $F$ we again have an approximating sequence $\{F_n\}$. If $F_n(x,u,w_n) \to F(x,u,w)$ for all $(x,u) \in\mathds{X}\times\mathds{U}$ and for any $w_n \to w$, then the transition kernel $\mathcal{T}_n$ corresponding to the model $F_n$ converges weakly to the one of $F$, $\mathcal{T}$: For any $g \in C_b(\mathds{X})$,
\begin{align*}
&\lim_{n \to \infty}\int g(x_1)\mathcal{T}_n(dx_1|x_0,u_0) =\lim_{n \to \infty}\int g(F_n(x_0,u_0,w)) \mu_n(dw)\\
& =\int g(F(x_0,u_0,w)) \mu(dw)= \int g(x_1)\mathcal{T}(dx_1|x_0,u_0). 
\end{align*}
In the analysis above we used a generalized dominated convergence result, Lemma \ref{langen}, to be presented later building on \cite[Theorem 3.5]{Langen81} and \cite[Theorem 3.5]{serfozo82}. 

\item[(v)] Let again $\{F_n\}$ denote an approximating sequence for $F$ and suppose now $F_{x_0,u_0,n}(\cdot):=F_n(x_0,u_0,\cdot): \mathds{W} \to \mathds{X}$ is invertible for all $x_0,u_0 \in \mathds{X}\times\mathds{U}$ and $F_{(x_0,u_0),n}^{-1}(\cdot)$ denotes the inverse for fixed $(x_0,u_0)$.  It is assumed that $F_{(x_0,u_0),n}^{-1}(x_1) \to F_{x_0,u_0}^{-1}(x_1)$ pointwise for all $(x_0,u_0)$. Suppose further that the noise process $w_t$ admits a continuous density $f_W(w)$. The Jacobian matrix, $\frac{\partial x_1}{\partial w}$, is the matrix whose components are the partial derivatives of $x_1$, i.e. with $x_1 \in \mathds{X}\subset \mathds{R}^m$ and $w\in\mathds{W}\subset\mathds{R}^m$, it is an $m\times m$ matrix with components $\frac{\partial (x_1)_i}{\partial w_j}$, $1\leq i,j\leq m$ . If the Jacobian matrix of derivatives $\frac{\partial x_1}{\partial w}(w)$ is continuous in $w$ and nonsingular for all $w$ then by the inverse function theorem of vector calculus (see \cite[Section 1.11]{hajek}), we have that the density of the state variables can be written as
\begin{align*}
&f_{X_1,n,(x_0,u_0)}(x_1)=f_W(F_{x_0,u_0,n}^{-1}(x_1))\big|\frac{\partial x_1}{\partial w}(F_{x_0,u_0,n}^{-1}(x_1))\big|^{-1}\\
&f_{X_1,(x_0,u_0)}(x_1)=f_W(F_{x_0,u_0}^{-1}(x_1))\big|\frac{\partial x_1}{\partial w}(F_{x_0,u_0}^{-1}(x_1))\big|^{-1}.
\end{align*}
With the above, $f_{X_1,n,(x_0,u_0)}(x_1) \to f_{X_1,(x_0,u_0)}(x_1)$ pointwise for all fixed $(x_0,u_0)$. Therefore, by Scheff\'e's Theorem, the corresponding kernels $\mathcal{T}_n(\cdot|x_0,u_0) \to \mathcal{T}(\cdot|x_0,u_0)$ in total variation for all $(x_0,u_0)$.

\item[(vi)] These studies will be used and analyzed in detail in Section \ref{empiricalLearning}, where data-driven stochastic control problems will be considered where estimated models are obtained through empirical measurements of the state action variables. 
\end{itemize}
%
%
\subsection{Summary of results and contributions}\label{contributions}
We now introduce the main assumptions that will be occasionally used for our technical results in the paper.

\begin{assumption}\label{weak_assmp}
\begin{itemize}
\item[(a)] The sequence of transition kernels $\mathcal{T}_n$ satisfies the following; $\{\mathcal{T}_n(\cdot|x_n,u_n),n\in\mathds{N}\}$ converges weakly to $\mathcal{T}(\cdot|x,u)$ for any sequence $\{x_n,u_n\}\subset\mathds{X}\times\mathds{U}$ and $x,u \in \mathds{X}\times\mathds{U}$ such that $(x_n,u_n)\to (x,u)$,
\item [(b)] The stochastic kernels $\mathcal{T}(\cdot|x,u)$ and $\{\mathcal{T}_n(\cdot|x,u)\}_n$ are weakly continuous in $(x,u)$,
\item[(c)] The stage-wise cost function $c(x,u)$ is non-negative, bounded and continuous on $\mathds{X} \times \mathds{U}$.
\item [(d)] $\mathds{U}$ is compact. 
\end{itemize}
\end{assumption}

\begin{assumption}\label{weak_assmpChannel}
\begin{itemize}
\item The observation channel $Q(\cdot|x)$ is continuous in total variation i.e., if $x_k \rightarrow x$, then $Q(\,\cdot\,|x_k) \rightarrow Q(\,\cdot\,|x)$ in total variation (only for partially observed models),
\end{itemize}
\end{assumption}

\begin{assumption}\label{set_assmp}
\begin{itemize}
\item[(a)] The sequence of transition kernels $\mathcal{T}_n$ satisfies the following; $\{\mathcal{T}_n(\cdot|x,u_n),n\in\mathds{N}\}$ converges setwise to $\mathcal{T}(\cdot|x,u)$ for any sequence $\{u_n\}\subset\mathds{U}$ and $x,u \in \mathds{X}\times\mathds{U}$ such that $u_n\to u$,
\item [(b)] The stochastic kernels $\mathcal{T}(\cdot|x,u)$ and $\{\mathcal{T}_n(\cdot|x,u)\}_n$ are setwise continuous in $u$,
\item[(c)] The stage-wise cost function $c(x,u)$ is non-negative, bounded and continuous on $\mathds{U}$.
\item [(d)] $\mathds{U}$ is compact. 
\end{itemize}
\end{assumption}

\begin{assumption}\label{tv_assmp}
\begin{itemize}
\item[(a)] The sequence of transition kernels $\mathcal{T}_n$ satisfies the following; $\|\mathcal{T}_n(\cdot|x,u_n)-\mathcal{T}(\cdot|x,u)\|_{TV}\to 0$  for any sequence $\{u_n\}\subset\mathds{U}$ and $x,u \in \mathds{X}\times\mathds{U}$ such that $u_n\to u$,
\item [(b)] The stochastic kernels $\mathcal{T}(\cdot|x,u)$ and $\{\mathcal{T}_n(\cdot|x,u)\}_n$ are continuous in total variation in $u$,
\item[(c)] The stage-wise cost function $c(x,u)$ is non-negative, bounded and continuous on $\mathds{U}$.
\item [(d)] $\mathds{U}$ is compact. 
\end{itemize}
\end{assumption}

In Section \ref{continuity} and \ref{robustness} we study continuity (Problem P1) and robustness (Problem P2) for partially observed models. In particular we show that 
\begin{itemize}
\item Continuity and robustness do not hold in general under weak convergence of kernels (Theorem \ref{weakneg}).
\item Under Assumptions \ref{weak_assmp} and \ref{weak_assmpChannel}, continuity and robustness hold (Theorem \ref{weak_cont}, Theorem \ref{weak_robust_kernel}).
\item Continuity and robustness do not hold in general under setwise convergence of the kernels (Theorem \ref{setwise_neg}).
\item Continuity and robustness do not hold in general under total variation convergence of the kernels (Example \ref{cont_conv_counter}).
\item Under Assumption \ref{tv_assmp},  continuity and robustness hold (Theorem \ref{TV_cont_thm}, Theorem \ref{TV_robust_kernel2}).
\end{itemize}

In Section \ref{fully}, we study continuity (Problem P1) and robustness (Problem P2) for fully observed models. In particular we show that 
\begin{itemize}
\item Continuity and robustness do not hold in general under weak convergence of kernels (Theorem \ref{weakneg_fully}, Example \ref{cont_conv_counter}).
\item Under Assumption \ref{weak_assmp}, continuity holds (Theorem \ref{fully_weak_suff_cont}), under Assumption \ref{weak_assmp}, robustness holds if the optimal policies for every initial point are identical (Theorem \ref{robust_fully}).
\item Continuity and robustness do not hold in general under setwise convergence of the kernels (Theorem \ref{notContSetwiseFull}, Theorem \ref{notRobustSetwiseFull}).
\item Under Assumption \ref{set_assmp}, continuity holds (Theorem \ref{fully_set_suff_cont}), under Assumption \ref{set_assmp}, robustness holds if the optimal policies for every initial point are identical (Theorem \ref{robust_fully_set}).
\item Continuity and robustness do not hold in general under total variation convergence of the kernels (Example \ref{cont_conv_counter}).
\item Under Assumption \ref{tv_assmp},  continuity and robustness hold (Subsection \ref{fully_TV}).
\end{itemize}

 Compared to the existing literature reviewed earlier, the above results use strictly more relaxed and refined convergence criteria to study robustness. In Section \ref{empiricalLearning}, these results will be applied to arrive at positive and negative implications on empirical consistency and data-driven learning in stochastic control since often system models are learned through empirical training data where typically weak convergence criterion applies (in an almost sure sense) but stronger convergence criteria do not.

\section{Continuity of Optimal Cost with respect to Convergence of Transition Kernels (Partially Observed Case)}\label{continuity}

In this section, we will study continuity of the optimal discounted cost under the convergence of transition kernels for partially observed models. 

\subsection{Weak convergence}
\subsubsection{Absence of continuity under weak convergence}
The following shows that the optimal cost may not be continuous under weak convergence of transition kernels.
\begin{theorem}\label{weakneg}
Let $\mathcal{T}_n \to \mathcal{T}$ weakly, then it is not necessarily true that $J^*_\beta(P,\mathcal{T}_n) \to J^*_\beta(P,\mathcal{T})$ even when the prior distributions are same, the measurement channel $Q$ is continuous in total variation and $c(x,u)$ is continuous and bounded on $\mathds{X}\times \mathds{U}$.
\end{theorem} 
\proof
We prove the result with a counterexample.
Let  $\mathds{X}=\mathds{U}=\mathds{Y}=[-1,1]$ and $c(x,u)=(x-u)^2$, the observation channel is chosen to be uniformly distributed over [-1,1], $Q \sim U([-1,1])$, 
the initial distributions of the state variable are chosen to be same as $P \sim \delta_1$ where $\delta_A(x):=\mathds{1}_{\{x \in A\}}$ for Borel $A$, and the transition kernels are:
\begin{align*}
 \mathcal{T}(\cdot|x,u)&=\delta_{-1}(x)[\frac{1}{2}\delta_1(\cdot)+ \frac{1}{2}\delta_{-1}(\cdot)]+ \delta_{1}(x)[\frac{1}{2}\delta_1(\cdot) + \frac{1}{2}\delta_{-1}(\cdot)] \\
& \quad \quad \quad \quad \quad \quad \quad+ (1-\delta_{-1}(x))(1-\delta_1(x))\delta_0(\cdot) \\
 \mathcal{T}_n(\cdot|x,u)&= \delta_{-1}(x)[\frac{1}{2}\delta_{(1-1/n)}(\cdot) + \frac{1}{2}\delta_{(-1 +1/n)}(\cdot)] + \delta_{1}(x)[\frac{1}{2}\delta_{(1 -1/n)}(\cdot) + \frac{1}{2}\delta_{(-1+1/n)}(\cdot)] \\
 & \quad \quad \quad \quad \quad \quad \quad+ (1-\delta_{-1}(x))(1-\delta_1(x))\delta_0(\cdot). 
\end{align*}
In other words, for $\mathcal{T}$:
\begin{align*}
&\text{If } x_t \in \{1,-1\} \text{ then} \quad x_{t+1}=1 \text{ or } -1 \\
& \text{else } x_{t+1}=0
\end{align*}
for $\mathcal{T}_n$
\begin{align*}
&\text{If } x_t \in \{1,-1\} \text{ then} \quad x_{t+1}=1-\frac{1}{n} \text{ or } -1+\frac{1}{n}\\
&\text{else } x_{t+1}=0
\end{align*}
independent of the control, where the events noted above with {\it or} are equally likely. It can be seen that $ \mathcal{T}_n \to  \mathcal{T}$ weakly according to Definition \ref{def_kernel}(i). Since the cost function is mean square error, control does not affect the dynamics and the channel is non-informative, the optimal policy is: 
\begin{align*}
\gamma_k^*(y_{[0,k]})=E[X_k]=&\begin{cases}0 \quad \text{ if } k>0\\
1  \quad \text{ if } k=0.\end{cases}
\end{align*}
Note that the cost function is continuous, and the measurement channel is continuous in total variation. The optimal discounted costs can be found as:
\begin{align*}
&J_{\beta}^*(P,\mathcal{T})=\sum_{k=1}^{\infty}E_P^{{\mathcal{T}}}[\beta^kX_k^2]=\sum_{k=1}^{\infty}\beta^k=\frac{\beta}{1-\beta}\\
&J_{\beta}^*(P,\mathcal{T}_n)=\sum_{k=1}^{\infty}E_P^{{\mathcal{T}_n}}[\beta^kX_k^2]=\beta[\frac{1}{2}(1-\frac{1}{n})^2+\frac{1}{2}(-1+\frac{1}{n})^2].
\end{align*}
Then we have $J_{\beta}^*(P, \mathcal{T}_n) \to \beta \neq \frac{\beta}{1-\beta}$
\endproof

\subsection{A sufficient condition for continuity under weak convergence} \label{partially_weak_cont}
In the following, we will establish and utilize some regularity properties for the optimal cost with respect to the convergence of transition kernels.

\begin{assumption}
\label{weak:as3}
\begin{itemize}
\item [(a)] The stochastic kernel $\mathcal{T}(\cdot|x,u)$ is weakly continuous in $(x,u)$.
\item [(b)] The observation channel $Q(\cdot|x)$ is continuous in total variation i.e., if $x_k \rightarrow x$, then $Q(\,\cdot\,|x_k) \rightarrow Q(\,\cdot\,|x)$ in total variation.
\item[(c)] The stage-wise cost function $c(x,u)$ is non-negative, bounded and continuous on $\mathds{X} \times \mathds{U}$
\item [(d)] $\mathds{U}$ is compact. 
\end{itemize}
\end{assumption}

It is a standard result that any partially observed Markov decision process (POMDP) can be reduced to a (completely observable) MDP, whose states are the posterior state distributions or {\it beliefs} of the observer; that is, the state at time $t$ is $Z_t(\,\cdot\,) := \sPr\{X_{t} \in \,\cdot\, | Y_0,\ldots,Y_t, U_0, \ldots, U_{t-1}\} \in \P(\sX)$. We call this equivalent MDP the belief-MDP \index{Belief-MDP}. The belief-MDP has state space $\sZ = \P(\sX)$ and action space $\sU$. Under the topology of weak convergence, since $\sX$ is a Borel space, $\sZ$ is metrizable with the Prokhorov metric which makes $\sZ$ into a Borel space \cite{Par67}. The transition probability $\eta$ of the belief-MDP can be constructed through non-linear filtering equations \cite[p. 334-335]{SYL2016near}. 
The one-stage cost function $c$ of the belief-MDP is given by $\tilde{c}(z,u) := \int_{\sX} c(x,u) z(dx)$. By \cite[Proposition 7.30]{bertsekas78}, the one stage cost function $\tilde{c}$ of the belief-MDP is continuous and bounded, that is in $C_b(\sZ\times\sU)$, under Assumption \ref{weak:as3}-(c). By \cite[Theorem 3.7, Example 4.1]{FeKaZg14} (see also \cite{KSY2019scl}), under Assumption \ref{weak:as3}, the stochastic kernel $\eta$ for belief-MDP is weakly continuous in $(z,u)$. For an MDP with weakly continuous transition probabilities and compact action spaces, it follows that an optimal control policy exists: This follows because the {\it discounted cost optimality operator} $T: C_b(\sZ) \to C_b(\sZ)$ (see e.g. \cite[Chapter 8.5]{hernandezlasserre1999further}):
\begin{align}
(T(f))(z) = \min_{u} ( \tilde{c}(z,u) + \beta E[f(z_1) | z_0=z, u_0=u]) \label{DCOE}
\end{align}
is a contraction from $C_b(\sZ)$ to itself under the supremum norm. As a result, there exists a fixed point, the value function, and an optimal control policy exists. In view of this existence result, in the following we will consider optimal policies. We note though that for the results which do not use the assumption, one may use $\epsilon$-optimal policies without affecting the results.




\begin{theorem}\label{same_policy_cont}
 Under Assumptions \ref{weak_assmp} and \ref{weak_assmpChannel},
\[\sup_{\gamma \in \Gamma}|J_\beta(P,\mathcal{T}_n,\gamma)-J_\beta(P,\mathcal{T},\gamma)|\to 0.\]
\end{theorem}
\proof
\begin{align*}
&\sup_{\gamma \in \Gamma}|J_\beta(P,\mathcal{T}_n,\gamma)-J_\beta(P,\mathcal{T},\gamma)| \\
&\qquad= \sup_{\gamma \in \Gamma}\bigg|\sum_{t=0}^{\infty}\beta^t\bigg(E_P^{{\mathcal{T}}}\Big[c\big(X_t,\gamma(Y_{[0,t]})\big)\Big]-E_P^{{\mathcal{T}_n}}\Big[c\big(X_t,\gamma(Y_{[0,t]})\big)\Big]\bigg)\bigg|\\
&\qquad \leq \sum_{t=0}^{\infty}\beta^t \sup_{\gamma \in \Gamma}\bigg|E_P^{{\mathcal{T}}}\Big[c\big(X_t,\gamma(Y_{[0,t]})\big)\Big]-E_P^{{\mathcal{T}_n}}\Big[c\big(X_t,\gamma(Y_{[0,t]})\big)\Big]\bigg|.
\end{align*}
Recall that an {\em admissible policy} $\gamma$ is a
sequence of control functions $\{\gamma_t,\, t\in \Zplus\}$. At the last step above, we make a slight abuse of notation; the $\sup$ at the first step is over all sequence of control functions $\{\gamma_t,\, t\in \Zplus\}$ whereas  the $\sup$  at the last step is over all sequence of control functions $\{\gamma_{t'}, \, t'\leq t\}$ but we will use the same notation, $\gamma$, in the rest of the proof. In Appendix \ref{app1}, we show the following for any $t\geq0$:
\begin{align}\label{time_t_to_0}
&\sup_{\gamma\in \Gamma}\Big|E_P^{{\mathcal{T}}}\bigg[c\big(X_t,\gamma(Y_{[0,t]})\big)\bigg]-E_P^{{\mathcal{T}_n}}\bigg[c\big(X_t,\gamma(Y_{[0,t]})\big)\bigg]\Big|\to0.
\end{align}
For any $\epsilon>0$, we choose a $K<\infty$ such that $\sum_{t=K+1}^{\infty}\beta^k2\|c\|_\infty \leq \epsilon/2$. For the chosen $K$,  we choose an $N < \infty$ such that
\begin{align*}
\sup_{\gamma \in \Gamma}\bigg|E_P^{{\mathcal{T}}}\Big[c\big(X_t,\gamma(Y_{[0,t]})\big)\Big]-E_P^{{\mathcal{T}_n}}\Big[c\big(X_t,\gamma(Y_{[0,t]})\big)\Big]\bigg|\leq \epsilon/2K
\end{align*}
 for all time stages $t\leq K$ and for all $n > N$. Thus, we have that $\sup_{\gamma \in \Gamma}\big|J_\beta(P,\mathcal{T}_n,\gamma)-J_\beta(P,\mathcal{T},\gamma) \to 0$ as $n \to \infty$.
\endproof

Now we give the main result of this section.
\begin{theorem}\label{weak_cont}
Suppose the conditions of Theorem \ref{same_policy_cont} hold. Then
\begin{align*}
\lim_{n\to\infty}|J_\beta^*(P,\mathcal{T}_n)-J_\beta^*(P,\mathcal{T})|=0.
\end{align*}
\end{theorem}
\proof
We start with the following bound,
\begin{align}\label{policy_exchange}
&|J_{\beta}^*(P, \mathcal{T}_n) - J_{\beta}^*(P, \mathcal{T})| \nonumber\\
&\qquad\qquad\leq \max\bigg(J_\beta(P,\mathcal{T}_n,\gamma^*)- J_{\beta}(P,\mathcal{T},\gamma^*),J_\beta(P,\mathcal{T},\gamma_n^*)- J_{\beta}(P,\mathcal{T}_n,\gamma_n^*)\bigg) 
\end{align}
where $\gamma^*$ and $\gamma_n^*$ are the optimal policies respectively for $ \mathcal{T}$ and $ \mathcal{T}_n$. Both terms go to $0$ by Theorem \ref{same_policy_cont}.
\endproof

\subsection{Absence of continuity under setwise convergence}
We now show that continuity of optimal costs may fail under the setwise convergence of transition kernels. Theorem \ref{notContSetwiseFull} in the next section establishes this result for fully observed models. As we note later, a fully observed system can be viewed as a partially observed system with the measurement being the state itself through (\ref{QMDPasPOMDP}), therefore, in view of space constraints, a separate proof will not be provided for the following result.
\begin{theorem}\label{setwise_neg}
Let $\mathcal{T}_n \to \mathcal{T}$ setwise. Then, it is not true in general that $J^*_\beta(P,\mathcal{T}_n)\to J^*_\beta(P,\mathcal{T})$, even when $\mathds{X}, \mathds{Y}$ and $\mathds{U}$ are compact and $c(x,u)$ is continuous and bounded in $\mathds{X}\times\mathds{U}$.
\end{theorem}

\subsection{Continuity under total variation} 
We have the following results.
%
\begin{theorem}\label{TV_cont_thm}
Under Assumption \ref{tv_assmp}
\[ J^*_\beta(P,\mathcal{T}_n) \to J^*_\beta(P,\mathcal{T}).\]
\end{theorem}
\proof
We start with the following bound,
\begin{align*}
&|J_{\beta}^*( \mathcal{T}_n) - J_{\beta}^*( \mathcal{T})| \leq \max\bigg(J_\beta( \mathcal{T}_n,\gamma^*)- J_{\beta}( \mathcal{T},\gamma^*),J_\beta( \mathcal{T},\gamma_n^*)- J_{\beta}( \mathcal{T}_n,\gamma_n^*)\bigg) 
\end{align*}
where $\gamma^*$ and $\gamma_n^*$ are the optimal policies respectively for $ \mathcal{T}$ and $ \mathcal{T}_n$.

We now study the following:
\begin{align*}
&\sup_{\gamma \in \Gamma}|J_\beta(P,\mathcal{T}_n,\gamma)-J_\beta(P,\mathcal{T},\gamma)|\\
&\qquad= \sup_{\gamma \in \Gamma}\bigg|\sum_{t=0}^{\infty}\beta^t\bigg(E_P^{{\mathcal{T}}}\Big[c\big(X_t,\gamma(Y_{[0,t]})\big)\Big]-E_P^{{\mathcal{T}_n}}\Big[c\big(X_t,\gamma(Y_{[0,t]})\big)\Big]\bigg)\bigg|\\
&\qquad \leq \sum_{t=0}^{\infty}\beta^t \sup_{\gamma \in \Gamma}\bigg|E_P^{{\mathcal{T}}}\Big[c\big(X_t,\gamma(Y_{[0,t]})\big)\Big]-E_P^{{\mathcal{T}_n}}\Big[c\big(X_t,\gamma(Y_{[0,t]})\big)\Big]\bigg|
\end{align*}

In Appendix \ref{app2} we show that for all $t<\infty$
\begin{align}\label{TV_time_t_to_0}
&\sup_{\gamma \in \Gamma}\bigg|E_P^{{\mathcal{T}}}\Big[c\big(X_t,\gamma(Y_{[0,t]})\big)\Big]-E_P^{{\mathcal{T}_n}}\Big[c\big(X_t,\gamma(Y_{[0,t]})\big)\Big]\bigg|\to 0.
\end{align}

For any $\epsilon>0$, we choose a $K<\infty$ such that $\sum_{t=K+1}^{\infty}\beta^t2\|c\|_\infty \leq \epsilon/2$. For the chosen $K$,  we choose an $N < \infty$ such that
\begin{align*}
\sup_{\gamma \in \Gamma}\bigg|E_P^{{\mathcal{T}}}\Big[c\big(X_t,\gamma(Y_{[0,t]})\big)\Big]-E_P^{{\mathcal{T}_n}}\Big[c\big(X_t,\gamma(Y_{[0,t]})\big)\Big]\leq \epsilon/2K
\end{align*}
 for all time stages $t\leq K$ and for all $n > N$. Therefore, we have that for any given $\epsilon>0$, for $n>N$ 
\begin{align}\label{TV_cont}
\sup_{\gamma \in \Gamma}\big|J_\beta(\mathcal{T}_n,\gamma)-J_\beta(\mathcal{T},\gamma)\big| < \epsilon.
\end{align}
 Thus, the result follows.
\endproof

We now present a result on the rate of convergence. For stochastic control problems, {\it strategic measures} are defined \cite{Schal} as the set of probability measures induced on the product spaces of the state and action pairs by admissible control policies: Given an initial distribution on the state, and a policy, one can uniquely define a probability measure on the infinite product space consistent with finite dimensional distributions, by the Ionescu Tulcea theorem \cite[Proposition C.10]{HernandezLermaMCP}. Now, define a strategic measure under a policy $\gamma^n= \{\gamma^n_0,\gamma^n_1, \cdots, \gamma^n_k,\cdots\}$ as a probability measure defined on ${\cal B}(\mathds{X} \times \mathds{Y} \times \mathds{U})^{\mathds{Z}_+}$ by:
\begin{align*}
&P^{\gamma^n}_{\mathcal{T}}(\dd(x_0,y_0,u_0),\dd(x_1,y_1,u_1),\cdots) \nonumber \\
& \quad = P(\dd x_0) Q(\dd y_0|x_0) 1_{\{\gamma^n(y_0) \in \dd u_0\}} \mathcal{T}(\dd x_1|x_0,u_0) Q(\dd y_1|x_1) 1_{\{\gamma^n(y_0,y_1) \in \dd u_1\}} \cdots
\end{align*}

Next, with uniformity in the total variation convergence, Theorem \ref{TV_cont_thm} is enhanced. 
\begin{theorem}\label{TV_bound_kernel}
If the cost function $c$ is bounded,
\begin{align*}
|J_{\beta}^*(P, \mathcal{T}_n) - J_{\beta}^*(P, \mathcal{T})| \leq \|c\|_\infty\frac{\beta}{(\beta -1)^2}  \sup_{x \in \mathds{X}, u \in \mathds{U}} \| \mathcal{T}_n(.|x,u)- \mathcal{T}(.|x,u)\|_{TV}.
\end{align*}
\end{theorem}

\proof
We start with the following bound as before,
\begin{align*}
&|J_{\beta}^*( \mathcal{T}_n) - J_{\beta}^*( \mathcal{T})| \leq \max\bigg(J_\beta( \mathcal{T}_n,\gamma^*)- J_{\beta}( \mathcal{T},\gamma^*),J_\beta( \mathcal{T},\gamma_n^*)- J_{\beta}( \mathcal{T}_n,\gamma_n^*)\bigg) 
\end{align*}
where $\gamma^*$ and $\gamma_n^*$ are the optimal policies respectively for $ \mathcal{T}$ and $ \mathcal{T}_n$.

Then, with $P^\gamma_{\mathcal{T}_n}$ and $P^\gamma_{\mathcal{T}}$ denoting the strategic measures for two chains with a policy $\gamma$ and kernels $\mathcal{T}_n$ and $\mathcal{T}$ , we have
\begin{align*}
&|J_{\beta}( \mathcal{T}_n,\gamma) - J_{\beta}( \mathcal{T},\gamma)|\\
 &\leq \sum_k \beta^k|\int c(x_k,\gamma(y_{[0,k]}))P^\gamma_{ \mathcal{T}_n}(dx_k,dy_{[0,k]}) - \int c(x_k,\gamma(y_{[0,k]}))P^\gamma_{ \mathcal{T}}(dx_k,dy_{[0,k]})| \\
&\leq \sum_k \beta^k \|c\|_\infty \|P^\gamma_{ \mathcal{T}_n}(d(x,y,u)_{[0,k]})-P^\gamma_{ \mathcal{T}}(d(x,y,u)_{[0,k]})\|_{TV}
\end{align*}

In Appendix \ref{app3} we establish the following relation:
\begin{align}\label{TV_sup_bound}
\|P^\gamma_{ \mathcal{T}_n}(d(x,y,u)_{[0,k]})-P^\gamma_{ \mathcal{T}}(d(x,y,u)_{[0,k]})\|_{TV} \leq k  \sup_{x \in \mathds{X}, u \in \mathds{U}} \| \mathcal{T}_n(.|x,u)- \mathcal{T}(.|x,u)\|_{TV}.
\end{align}
 Using this bound, we will have
\begin{align*}
\|J_{\beta}^*( \mathcal{T}_n) - J_{\beta}^*( \mathcal{T})| &\leq \sum_k \beta^k \|c\|_\infty  k  \sup_{x \in \mathds{X}, u \in \mathds{U}} \| \mathcal{T}(.|x,u)- \mathcal{T}_n(.|x,u)\|_{TV}\\
&= \|c\|_\infty\frac{\beta}{(\beta -1)^2}  \sup_{x \in \mathds{X}, u \in \mathds{U}} \| \mathcal{T}(.|x,u)- \mathcal{T}_n(.|x,u)\|_{TV}.
\end{align*}
$\hfill$ \endproof

\section{Robustness to Incorrect Transition Kernels (Partially Observed Case)}\label{robustness}
Here, we consider the robustness problem {\bf P2}: Suppose we design an optimal policy, $\gamma_n^*$, for a transition kernel, $\mathcal{T}_n$, assuming it is the correct model and apply the policy to the true model whose transition kernel is $\mathcal{T}$. We study the robustness of the sub-optimal policy $\gamma_n^*$. 
\subsection{Total variation} \hfill\\

\begin{theorem}\label{TV_robust_kernel}
Suppose the stage-wise cost function $c(x,u)$ is bounded in $\mathds{X}\times \mathds{U}$, then
\begin{align*}
|J_\beta(P, \mathcal{T},\gamma_n^*)-J_\beta^*(P, \mathcal{T})|\leq 2 \|c\|_\infty\frac{\beta}{(\beta -1)^2}  \sup_{x \in \mathds{X}, u \in \mathds{U}} \| \mathcal{T}(.|x,u)- \mathcal{T}_n(.|x,u)\|_{TV} 
\end{align*} 
for a fixed prior distribution $P \in \mathcal{P}(\mathds{X})$, where $\gamma_n^*$ is the optimal policy designed for the transition kernel $\mathcal{T}_n$.
\end{theorem}


\proof
We begin with the following,
\begin{align*}
&|J_\beta(\mathcal{T},\gamma_n^*)-J_\beta^*(\mathcal{T})| \leq |J_\beta(\mathcal{T},\gamma_n^*)-J_\beta(\mathcal{T}_n,\gamma_n^*)|+|J_\beta(\mathcal{T}_n,\gamma_n^*)-J_\beta(\mathcal{T},\gamma^*)|
\end{align*}
The second term is bounded using Theorem \ref{TV_bound_kernel}. For the first term, we use the proof of Theorem \ref{TV_bound_kernel} where we showed that for any $\gamma \in \Gamma$
\[ |J_\beta(\mathcal{T},\gamma)-J_\beta(\mathcal{T}_n,\gamma)| \leq \|c\|_\infty\frac{\beta}{(\beta -1)^2}  \sup_{x \in \mathds{X}, u \in \mathds{U}} \| \mathcal{T}(.|x,u)- \mathcal{T}_n(.|x,u)\|_{TV}.\]
Thus, the result follows.
\endproof

The next theorem gives an asymptotic robustness result.
\begin{theorem}\label{TV_robust_kernel2}
Under Assumption \ref{tv_assmp}
\[|J_\beta(P,\mathcal{T},\gamma_n^*)-J_\beta^*(P,\mathcal{T})| \to 0\]
where $\gamma_n^*$ is the optimal policy designed for the kernel $\mathcal{T}_n$.
\end{theorem}
\proof
We write the following;
\begin{align*}
&|J_\beta(P,\mathcal{T},\gamma_n^*)-J_\beta^*(P,\mathcal{T})|\leq|J_\beta(P,\mathcal{T},\gamma_n^*)-J_\beta^*(P,\mathcal{T}_n)|+|J_\beta^*(P,\mathcal{T}_n)-J_\beta^*(P,\mathcal{T}).
\end{align*}
The second term goes to 0 by Theorem \ref{TV_cont_thm} and the first term goes to 0 using (\ref{TV_cont}) again from the proof of Theorem \ref{TV_cont_thm}.
\endproof

\subsection{Setwise convergence}
Theorem \ref{notRobustSetwiseFull} in the next section establishes the lack of robustness under setwise convergence of kernels. As we note later, a fully observed system can be viewed as a partially observed system with the measurement being the state itself, see (\ref{QMDPasPOMDP}).

\subsection{Weak convergence}\hfill\\
\begin{theorem}\label{weak_robust_kernel}
Under Assumptions \ref{weak_assmp} and \ref{weak_assmpChannel}, $|J_\beta(\mathcal{T},\gamma_n^*)-J_\beta^*(\mathcal{T})| \to 0$,
where $\gamma_n^*$ is the optimal policy designed for the transition kernel $\mathcal{T}_n$.
\end{theorem}
\proof
We write
\begin{align*}
&|J_\beta(\mathcal{T},\gamma_n^*)-J_\beta^*(\mathcal{T})| \leq |J_\beta(\mathcal{T},\gamma_n^*)-J_\beta(\mathcal{T}_n,\gamma_n^*)|+|J_\beta(\mathcal{T}_n,\gamma_n^*)-J_\beta(\mathcal{T},\gamma^*)|.
\end{align*}
The first term goes to 0 by Theorem \ref{same_policy_cont}. For the second term we use Theorem \ref{weak_cont}.
\endproof

\begin{remark}
In this paper we study the case where the channel is known to the controller; that is, the true channel model $Q$ is available to the controller. For the case where this is no longer true, the following analysis can be made. If the transition kernel $\mathcal{T}$ and the channel $Q$ are not known, the controller would have an approximating sequence $\mathcal{T}_nQ_n(x_{t+1},y_{t+1}\in \cdot\times \cdot|x_t,u_t) \in \P(\mathds{X}\times \mathds{Y})$ for the true joint measure $\mathcal{T}Q(x_{t+1},y_{t+1}\in \cdot\times \cdot|x_t,u_t) \in \P(\mathds{X}\times\mathds{Y})$ for all $(x_t,u_t)$. The question then becomes analyzing the convergence of $\mathcal{T}_nQ_n \to \mathcal{T}Q$. Due to space constraints, we do not present explicit results on this problem, however, we note that in \cite{yuksel12:siam}, a similar joint convergence is studied for convergence of measurement channels and fixed model/prior distributions. The reader can refer to \cite[Lemma 2.2]{yuksel12:siam} for an analysis on the convergence of $PQ_n \to PQ$, for a single stage problem and for a multi-stage problem, the following can be considered: With
\begin{align*}
|J_\beta^*(\mathcal{T}_nQ_n)- J^*_\beta(\mathcal{T}Q)|\leq |J_\beta^*(\mathcal{T}_nQ_n)-J_\beta^*(\mathcal{T}Q_n)|+|J_\beta^*(\mathcal{T}Q_n)-J_\beta^*(\mathcal{T}Q)|,
\end{align*}
\cite[Theorem 6.2]{yuksel12:siam} presents sufficient conditions to guarantee the convergence of the second term above. For the first term, the total variation convergence results in this paper provide an analysis on the uniform convergence over a class of channels, thus establishing positive results on continuity under the joint convergence of both transition kernels and measurement channels. 
\end{remark}

\section{Continuity and Robustness in the Fully Observed Case}\label{fully}

In this section, we consider the fully observed case where the controller has direct access to the state variables. We present the results for this case separately, since here we cannot utilize the regularity properties of measurement channels which allows for stronger continuity and robustness results. Similar to the discussions related to (\ref{DCOE}) that is as the operator defined in $(\ref{DCOE})$ is a contraction and as it admits a fixed point (value function), under measurable selection conditions due to weak or strong (setwise) continuity of transition kernels \cite[Section 3.3]{HernandezLermaMCP}, for infinite horizon discounted cost problems, optimal policies can be selected from those which are stationary and deterministic. Therefore we will restrict the policies to be stationary and deterministic so that $U_t=\gamma(X_t)$ for some measurable function $\gamma$. Notice also that fully observed models can be viewed as partially observed with the measurement channel thought to be
\begin{align}
Q(\cdot|x)=\delta_x(\cdot),\label{QMDPasPOMDP}
\end{align}
which is only weakly continuous, thus it does not satisfy Assumption \ref{weak_assmpChannel}.

\subsection{Weak convergence}

\subsubsection{Absence of continuity under weak convergence}
\hfill

We start with a negative result. 
\begin{theorem}\label{weakneg_fully}
For $\mathcal{T}_n \to \mathcal{T}$ weakly, it is not necessarily true that $J^*_\beta(\mathcal{T}_n) \to J^*_\beta(\mathcal{T})$ even when the prior distributions are same and $c(x,u)$ is continuous and bounded in $\mathds{X}\times \mathds{U}$.
\end{theorem} 
\proof
We prove the result with a counterexample, similar to the model used in the proof of Theorem \ref{weakneg}
Let  $\mathds{X}=[-1,1]$, $\mathds{U}=\{-1,1\}$ and $c(x,u)=(x-u)^2$, 
the initial distributions are given by $P \sim \delta_1$ that is $X_0=1$ and the transition kernels are
\begin{align*}
 \mathcal{T}(\cdot|x,u)&=\delta_{-1}(x)[\frac{1}{2}\delta_1(\cdot)+ \frac{1}{2}\delta_{-1}(\cdot)]+ \delta_{1}(x)[\frac{1}{2}\delta_1(\cdot) + \frac{1}{2}\delta_{-1}(\cdot)] \\
 & \quad \quad \quad \quad \quad \quad \quad \quad+ (1-\delta_{-1}(x))(1-\delta_1(x))\delta_0(\cdot) \\
 \mathcal{T}_n(\cdot|x,u)&= \delta_{-1}(x)[\frac{1}{2}\delta_{(1-1/n)}(\cdot) + \frac{1}{2}\delta_{(-1 +1/n)}(\cdot)] + \delta_{1}(x)[\frac{1}{2}\delta_{(1 -1/n)}(\cdot) + \frac{1}{2}\delta_{(-1+1/n)}(\cdot)] \\
 & \quad \quad \quad \quad \quad \quad \quad \quad+ (1-\delta_{-1}(x))(1-\delta_1(x))\delta_0(\cdot) .
\end{align*}
It can be seen that $ \mathcal{T}_n \to  \mathcal{T}$ weakly according to Definition \ref{def_kernel}(i). Under this setup we can calculate the optimal costs as follows;
\begin{align*}
J_{\beta}^*(\mathcal{T}_n)= \frac{1}{n^2} + \sum_{k=2}^\infty \beta^k = \frac{1}{n^2} + \frac{\beta^2}{1-\beta},
\end{align*}
and $J_{\beta}^*(\mathcal{T})=0$. Thus, continuity does not hold.
\endproof

We now present another counter example emphasizing the importance of continuous convergence in the actions. 
The following counter example shows that without the continuous convergence and regularity assumptions on the kernel $\mathcal{T}$, continuity fails even when $\mathcal{T}_n(\cdot|x,u)\to \mathcal{T}(\cdot|x,u)$ pointwise (for $x,u$) in total variation (also setwise and weakly) and even when the cost function $c(x,u)$ is continuous and bounded. Notice that this example also holds for setwise and weak convergence.

\begin{exmp}\label{cont_conv_counter}
Assume that the kernels are given by 
\begin{align*}
&\mathcal{T}_n(\cdot|x,u)\sim U([u^n,1+u^n])\\
&  \mathcal{T}(\cdot|x,u)\sim
      \begin{cases}
        U([0,1]) & \text{if } u\neq 1\\
        U([1,2])& \text{if } u=1
      \end{cases},
\end{align*}
where $\mathds{U} = [0,1]$ and $\mathds{X} = \mathds{R}$. We note first that $\mathcal{T}_n(\cdot|x,u)\to \mathcal{T}(\cdot|x,u)$ in total variation for every fixed $x$ and $u$.

The cost function is in the following form

\begin{align*}
c(x,u)=
 \begin{cases}
        2 & \text{if }  x\leq \frac{1}{e}\\
	2-\frac{x-\frac{1}{e}}{0.1}& \text{if } \frac{1}{e}< x\leq 0.1+\frac{1}{e}\\
        1& \text{if } 0.1+\frac{1}{e}< x\leq 1+\frac{1}{e}-0.1\\
	2-\frac{1+\frac{1}{e}-x}{0.1}&\text{if } 1+\frac{1}{e}-0.1< x\leq 1+\frac{1}{e}\\
	2& \text{if } 1+\frac{1}{e}< x 
      \end{cases}.
\end{align*}
Notice that $c(x,u)$ is a continuous function. 

With this setup, $\gamma^*(x)=0$ is an optimal policy for $\mathcal{T}$ since on $[0,1]$ interval the induced cost is less than the cost induced on $[1,2]$ interval. The cost under this policy is
\begin{align*}
 J_\beta^*(\mathcal{T})= \sum_{t=0}^\infty\beta^t\bigg(2\times\frac{1}{e}+\frac{0.3}{2}+0.9-\frac{1}{e}\bigg)=\frac{1}{1-\beta}\bigg(1.05+\frac{1}{e}\bigg).
\end{align*}

For $\mathcal{T}_n$, $\gamma_n^*(x)=e^{-\frac{1}{n}}$ is an optimal policy for every $n$ as $e^{-\frac{1}{n}\times n}=\frac{1}{e}$ and thus the state is distributed between $ \frac{1}{e}< x\leq 1+\frac{1}{e}$ in which interval the cost is the least. 
Hence, we can write
\begin{align*}
\lim_{n\to\infty} J_\beta(\mathcal{T}_n,\gamma_n^*)=\sum_{t=0}^\infty\beta^t\bigg(0.3+1-0.2\bigg)=\frac{1.1}{1-\beta}\neq \frac{1}{1-\beta}\bigg(1.05+\frac{1}{e}\bigg)= J_\beta^*(\mathcal{T}).
\end{align*}
$\hfill\diamond$


\end{exmp}

\subsubsection{A sufficient condition for continuity under weak convergence}
We will now establish that if the kernels and the model components have some further regularity, continuity does hold 


 The assumptions of the following result are same as the assumptions for the partially observed case (Theorem \ref{weak_cont}) except for the assumption on the measurement channel $Q$. For clarity, we present the assumptions here separately since for fully observed models we do not need a measurement channel in the analysis.
\begin{theorem}\label{fully_weak_suff_cont}
Under Assumption \ref{weak_assmp}, $J_\beta(\mathcal{T}_n,\gamma_n^*)\to J_\beta(\mathcal{T},\gamma^*)$, for any initial state $x_0$, as $n\to \infty$.
\end{theorem}
\proof 

We build on the proof of \cite[Proposition 3.10]{saldi2016markov}. We will use the successive approximations for an inductive argument.

Recall {\it discounted cost optimality operator} $T: C_b(\sZ) \to C_b(\sZ)$ from (\ref{DCOE})
\begin{align*}
(T(v))(x) = \inf_{u} ( c(x,u) + \beta E[v(x_1) | x_0=x, u_0=u]) 
\end{align*}
which is a contraction from $C_b(\sX)$ to itself under the supremum norm and has a fixed point, the value function.

 For the kernel $\mathcal{T}$, we will denote the approximation functions by
\[v^k(x)=T(v^{k-1})(x)\]
 and for the kernel $\mathcal{T}_n$  we will use $v^k_n(x)$ to denote the approximation functions, notice that the operator $T$ also depends on $n$ for the model $\mathcal{T}_n$ but we will continue using it as $T$ in what follows. 

We wish to show that the approximation functions for $\mathcal{T}_n$ continuously converge to the ones for $\mathcal{T}$. Then, for the first step of the induction we have
\begin{align*}
v^1(x)= c(x,u^*)\quad v_n^1(x_n)= c(x_n,u_n^*)
\end{align*}
thus we can write,
\begin{align*}
&|v^1(x)-v_n^1(x_n)|\leq \sup_{u \in \mathds{U}}\big|c(x,u)-c(x_n,u)\big|
\end{align*}
since $c\in C_b(\mathds{X}\times\mathds{U})$  and the action space, $\mathds{U}$, is compact, the first step of the induction holds, i.e. $\lim_{n\to\infty}|v^1(x)-v_n^1(x_n)|=0$. 

For the $k^{th}$ step we have,
\begin{align*}
&v^k(x)=T(v^{k-1})(x)=\inf_{u}\big[c(x,u)+\beta\int_{\mathds{X}}v^{k-1}(x^1)\mathcal{T}(dx^1|x,u)\big]\\
&v_n^k(x_n)=T(v_n^{k-1})(x_n)=\inf_{u}\big[c(x_n,u)+\beta\int_{\mathds{X}}v_n^{k-1}(x^1)\mathcal{T}_n(dx^1|x_n,u)\big].
\end{align*}
Note that the assumptions of the theorem satisfy the measurable selection criteria and hence we can choose minimizing selectors (\cite[Section 3.3]{HernandezLermaMCP}). If we denote the selectors by $u^*$ and $u_n^*$, we can write
\begin{align*}
&|v^k(x)-v_n^k(x_n)| \leq \\
& \max\Big(\bigg[|c(x,u^*)-c(x_n,u^*)|+\beta|\int_{\mathds{X}}v^{k-1}(x^1)\mathcal{T}(dx^1|x,u^*)-\int_{\mathds{X}}v_n^{k-1}(x^1)\mathcal{T}_n(dx^1|x_n,u^*)|\bigg],\\
&\bigg[|c(x,u_n^*)-c(x_n,u_n^*)|+\beta|\int_{\mathds{X}}v^{k-1}(x^1)\mathcal{T}(dx^1|x,u_n^*)-\int_{\mathds{X}}v_n^{k-1}(x^1)\mathcal{T}_n(dx^1|x_n,u_n^*)|\bigg]\Big).
\end{align*}
Hence, we can write
\begin{align}\label{action_exchange}
&\quad|v^k(x)-v_n^k(x_n)| \\
&\leq\sup_{u \in \mathds{U}}\bigg[|c(x,u)-c(x_n,u)|+\beta|\int_{\mathds{X}}v^{k-1}(x^1)\mathcal{T}(dx^1|x,u)-\int_{\mathds{X}}v_n^{k-1}(x^1)\mathcal{T}_n(dx^1|x_n,u)|\bigg]\nonumber
\end{align}
the first term goes to 0 as $c(x,u)$ is continuous in $x$ uniformly over all $u \in \mathds{U}$. For the second term we write,
\begin{align*}
&\sup_{u\in \mathds{U}}|\int_{\mathds{X}}v^{k-1}(x^1)\mathcal{T}(dx^1|x,u)-\int_{\mathds{X}}v_n^{k-1}(x^1)\mathcal{T}_n(dx^1|x_n,u)|\\
&\leq\sup_{u \in \mathds{U}} |\int_{\mathds{X}}\big(v^{k-1}(x^1)-v_n^{k-1}(x^1)\big)\mathcal{T}_n(dx^1|x_n,u)|\\
&\quad+\sup_{u \in \mathds{U}} |\int_{\mathds{X}}v^{k-1}(x^1)\mathcal{T}(dx^1|x,u)-\int_{\mathds{X}}v^{k-1}(x^1)\mathcal{T}_n(dx^1|x_n,u)|
\end{align*}
for the first term, by the induction argument for any $x^1_n \to x^1$, $\big|v^{k-1}(x^1)-v_n^{k-1}(x^1_n)\big| \to 0$ (i.e., we have continuous convergence). We also have that $\mathcal{T}_n(\cdot|x_n,u) \to \mathcal{T}(\cdot|x,u)$ weakly uniformly over $u \in \mathds{U}$ as $\mathds{U}$ is compact. Therefore, using Lemma \ref{langen} the first term goes to 0. For the second term we again use that $\mathcal{T}_n(\cdot|x_n,u)$ converges weakly to $\mathcal{T}(\cdot|x,u)$  uniformly over $u \in \mathds{U}$. With an almost identical induction argument it can also be shown that $v^{k-1}(x^1)$ is continuous in $x^1$, thus the second term also goes to 0.

So far, we have showed that for any $k\in\mathds{N}$, $\lim_{n\to \infty}\big|v^k_n(x_n)-v^k(x)\big|=0$ for any $x_n \to x$, in particular it is also true that  $\lim_{n\to \infty}\big|v^k_n(x)-v^k(x)\big|=0$ for any $x$. 

As we have stated earlier, it can be shown that the approximation operator, $T$ is a contractive operator under supremum norm with modulus $\beta$ and it converges to a fixed point which is the value function. Thus, we have
\begin{align}\label{contrac}
&\big|J_\beta(\mathcal{T},\gamma^*)-v^k(x)\big|\leq \|c\|_\infty\frac{\beta^k}{1-\beta},\quad\big|J^*_\beta(\mathcal{T}_n,\gamma_n^*)-v^k_n(x)\big|\leq \|c\|_\infty\frac{\beta^k}{1-\beta}.
\end{align}

Combining the results,
\begin{align*}
|J_\beta(\mathcal{T}_n,\gamma_n^*)-|J_\beta(\mathcal{T},\gamma^*)| \leq |J_\beta(\mathcal{T}_n,\gamma_n^*)-v^k_n(x)|+|v^k_n(x)-v^k(x)|+|J_\beta(\mathcal{T},\gamma^*)-v^k(x)|.
\end{align*}
Note that the first and the last term can be made arbitrarily small since (\ref{contrac}) holds for all $k\in\mathds{N}$; the second term goes to $0$ with an inductive argument for all $k \in \mathds{N}$.
\endproof

\subsubsection{Absence of robustness under weak convergence}\label{notRobustWeak}
The following result shows that the conditions that satisfy the continuity are not sufficient for robustness in the fully observed models.
\begin{theorem}\label{robust_counter_fully}
Suppose $\mathcal{T}_n(\cdot|x_n,u_n)\to\mathcal{T}(\cdot|x,u)$ weakly for every $x\in\mathds{X}$ and $u\in\mathds{U}$ and $(x_n,u_n)\to (x,u)$,  then it is not true in general that $J_\beta(\mathcal{T},\gamma_n^*)\to J_\beta(\mathcal{T},\gamma^*)$, even when $\mathds{X}$ and $\mathds{U}$ are compact and $c(x,u)$ is continuous and bounded in $\mathds{X}\times\mathds{U}$.
\end{theorem}
\proof
We prove the result with a counter example. Take $\mathds{X}=[0,2]$ and $\mathds{U}=\{0,1,2\}$.

 Suppose the kernels are given in the following form for $n\geq1$:
\begin{align*}
\mathcal{T}_n(\cdot|x,u)=&\delta_{1+1/n}(\cdot)\mathds{1}_{\{x\geq1+1/n,u=1\}}+ \delta_{1-1/n}(\cdot)\mathds{1}_{\{x\geq1+1/n,u=0\}}+\delta_{1}(\cdot)\mathds{1}_{\{x\geq1+1/n,u=2\}}\\
&+\delta_{1-1/n}\mathds{1}_{\{x\leq1-1/n,u=1\}}+ \delta_{1+1/n}\mathds{1}_{\{x\leq1-1/n,u=0\}}+\delta_{1}\mathds{1}_{\{x\leq1-1/n,u=2\}}\\
& +\delta_{1}\mathds{1}_{\{1-1/n< x<1+1/n\}}\\
\mathcal{T}(\cdot|x,u)=&\delta_{1}(\cdot).
\end{align*}
The cost function is given by:
\begin{align*}
c(x,u)=
\begin{cases}
(x-1)\mathds{1}_{x\geq1} + 0\mathds{1}_{x<1}  &\text{if } u=0,1\\
3 &\text{if } u=2.
\end{cases}
\end{align*}
With this setup, an optimal policy for $\mathcal{T}_n$ when the initial state is $x=0$ is given by;
\begin{align*}
\gamma_n^*(x)=
\begin{cases}
1  &\text{if } x\leq 1-1/n\\
0  &\text{if } x\geq 1+1/n\\
2  &\text{otherwise}.
\end{cases}
\end{align*}
When the initial state is $0$, the cost under this policy is $J_\beta(\mathcal{T}_n,\gamma_n^*)=0$, therefore the policy $\gamma_n^*$ is indeed optimal for $\mathcal{T}_n$. An  optimal policy for $\mathcal{T}$ is given by $\gamma^*(x)=1$. Thus, the discounted cost values can be calculated as:
\begin{align*}
J_\beta(\mathcal{T},\gamma_n^*)=&\sum_{t=0}^{\infty}\beta^tE[c(X_t,\gamma_n^*(X_t))]=\sum_{t=0}^{\infty}\beta^tc(1,\gamma_n^*(1))=\sum_{t=0}^{\infty}\beta^t3=\frac{3}{1-\beta}\\
J_\beta(\mathcal{T},\gamma^*)=&0.
\end{align*}
\endproof

\subsection{A sufficient condition for robustness under weak convergence}

We now present a result that establishes robustness if the optimal policies for every initial point are identical. That is, for every $n$, $\gamma_n^*$ is optimal for every $x_0\in\mathds{X}$ (under the model $\mathcal{T}_n$). Notice that in the counter example used for Theorem \ref{robust_counter_fully}, $\gamma_n^*$ is not optimal if the initial point is between $1-1/n$ and $1+1/n$. A sufficient condition for this property is that $\gamma_n^*$ solves the discounted cost optimality equation (DCOE) for every initial point.

A policy $\gamma^*\in\Gamma$ solves the discounted cost optimality equation and is optimal if it satisfies
\begin{align*}
J_\beta^*(\mathcal{T},x)=c(x,\gamma^*(x))+\beta \int J_\beta^*(\mathcal{T},x_1)\mathcal{T}(dx_1|x,\gamma^*(x)).
\end{align*}
Thus, a policy is optimal for every initial points if it satisfies the DCOE for all initial points $x\in\mathds{X}$.

\begin{theorem}\label{robust_fully}
Under Assumption \ref{weak_assmp}, $J_\beta(\mathcal{T},\gamma_n^*)\to J_\beta(\mathcal{T},\gamma^*)$ for any initial point $x_0$ if $\gamma_n^*$ is optimal for any initial point for the kernel $\mathcal{T}_n$. 
\end{theorem}

\proof

Define the following operator for $\gamma_n^*$, an optimal policy for $\mathcal{T}_n$,  
\begin{align}\label{operat}
(T_n(v))(x_0) =   c(x_0,\gamma_n^*(x_0)) + \beta \int v(x_1) \mathcal{T}(dx_1| x_0, \gamma_n^*(x_0))
\end{align}
which is a contraction from $C_b(\sX)$ to itself under the supremum norm with modulus $\beta$ and has a unique fixed point. One can show that the fixed point is $J_\beta(\mathcal{T},\gamma_n^*)$.

In the Appendix \ref{tech_operator_conv} we show that 
\begin{align}\label{k_conv}
T_n^k(J_\beta^*(\mathcal{T}))(x_n)\to J_\beta^*(\mathcal{T},x))
\end{align}
for any fixed $k<\infty$ as $n\to\infty$ for some $x_n\to x$ where $T_n^k$ denotes the operator $T_n$ applied $k$ consecutive times.

Our next claim is that $T_n^k(J_\beta^*(\mathcal{T}))(x)\to J_\beta(\mathcal{T},\gamma_n^*,x)$ as $k\to\infty$. This is true as $T_n$ is a contraction with modulus $\beta$ and $J_\beta(\mathcal{T},\gamma_n^*)$ is its unique fixed point. Thus $T_n^k(J_\beta^*(\mathcal{T}))(x)\to J_\beta(\mathcal{T},\gamma_n^*,x)$ as $k\to\infty$ and this convergence is uniform over $n$ as the contraction rate $\beta$ does not depend on $n$.

Now, we write:
\begin{align*}
J_\beta(\mathcal{T},\gamma_n^*)-J_\beta^*(\mathcal{T})\leq |J_\beta(\mathcal{T},\gamma_n^*)-T_n^k(J_\beta^*(\mathcal{T}))|+|T_n^k(J_\beta^*(\mathcal{T}))-J_\beta^*(\mathcal{T})|.
\end{align*}
We can make the first term arbitrarily small by choosing $k$ large enough uniformly over $n$. For the chosen $k$, the second term goes to $0$ as $n\to\infty$ by (\ref{k_conv}).

\endproof

Some remarks are in order.

\begin{remark}
For the partially observed case, the proof approach we use makes use of policy exchange (e.g. (\ref{policy_exchange})) and for this approach total variation continuity of channel $Q(\cdot|x)$ is a key step to deal with the uniform convergence over policies. As we stated before, the channel for fully observed models can be considered in the form of (\ref{QMDPasPOMDP}) which is only weakly continuous and not continuous in total variation. Thus, it may lead to negative results as in Theorem \ref{robust_counter_fully}.
However, for the fully observed models we can reach continuity and robustness (Theorem \ref{fully_weak_suff_cont}, Theorem \ref{robust_fully}) using value iteration approach. With this approach, instead of exchanging policies and analyzing uniform convergence over all policies, we can exchange control actions (e.g. (\ref{action_exchange})) and analyze uniform convergence over the action space $\mathds{U}$ by using the discounted optimality operator (\ref{DCOE}). Hence, we are only able to show convergence over optimal policies for the fully observed case, i.e. $J_\beta(\mathcal{T}_n,\gamma_n^*)\to J_\beta(\mathcal{T},\gamma^*)$ or $J_\beta(\mathcal{T},\gamma_n^*)\to J_\beta(\mathcal{T},\gamma^*)$ where $\gamma_n^*$ and $\gamma^*$ are optimal policies. Whereas, for partially observed models, regularity of the channel allows us to show convergence over any sequence of policies, i.e. $\sup_{\gamma\in\Gamma}|J_\beta(\mathcal{T}_n,Q,\gamma)-J_\beta(\mathcal{T},Q,\gamma)|\to 0$.
\end{remark}

\begin{remark}
As we have discussed in subsection \ref{partially_weak_cont}, a partially observed model can be reduced to a fully observed process where the state process (beliefs) becomes probability measure valued. Consider the partially observed models with transition kernels $\mathcal{T}_n$ and $\mathcal{T}$ (with a channel $Q$) and their corresponding fully observed transition kernels $\eta_n$ and $\eta$: following the discussions and techniques in \cite{FeKaZg_19} and \cite{KSY2019scl}, one can show that $\eta_n$ and $\eta$ satisfy the conditions of Theorem \ref{robust_fully} and Theorem \ref{fully_weak_suff_cont} that is $\eta_n(\cdot|z_n,u_n)\to \eta(\cdot|z,u)$ for any $(z_n,u_n)\to (z,u)$ under the following set of assumptions
\begin{itemize}
\item $\mathcal{T}_n(\cdot|x_n,u_n)\to \mathcal{T}(\cdot|x,u)$ for any $(x_n,u_n)\to (x,u)$,
\item $Q(\cdot|x)$ is continuous on total variation in $x$.
\end{itemize}
We remark that these conditions also agree with the conditions presented for continuity and robustness of the partially observed models (Theorem \ref{weak_cont} and Theorem \ref{weak_robust_kernel}).
\end{remark}

\begin{remark}\label{equ_remark} It can be shown that if we restrict the set of policies to an equicontinuous family of functions, robustness can also be achieved: Under the conditions of Theorem \ref{fully_weak_suff_cont}, in this case, $|J_\beta(\mathcal{T},\gamma_n^*)- J_\beta^*(\mathcal{T})|\to 0$. A short proof for this result can be found in Appendix \ref{equ_fully}.
\end{remark}

\subsection{Setwise convergence}
\subsubsection{Absence of continuity under setwise convergence}
We give a negative result similar to Theorem \ref{setwise_neg}.
\begin{theorem}\label{notContSetwiseFull}
Let $\mathcal{T}_n \to \mathcal{T}$ setwise then it is not necessarily true that $J^*_\beta(\mathcal{T}_n) \to J^*_\beta(\mathcal{T})$ even when $c(x,u)$ is continuous and bounded in $\mathds{X}\times \mathds{U}$.
\end{theorem}
\proof
See Example \ref{cont_conv_counter}. 
\endproof

\subsubsection{A sufficient condition for continuity under setwise convergence}
\hfill
\begin{theorem}\label{fully_set_suff_cont}
Under Assumption \ref{set_assmp} $J_\beta(\mathcal{T}_n,\gamma_n^*)\to J_\beta(\mathcal{T},\gamma^*)$, for any initial state $x_0$, as $n\to \infty$.
\end{theorem}
\begin{proof}
 We use the same value iteration technique that we used to prove Theorem \ref{fully_weak_suff_cont}.

We wish to show that the approximation functions for $\mathcal{T}_n$ converge pointwise to the ones for $\mathcal{T}$. Then, for the first step of the induction we have
\begin{align*}
v^1(x)= \inf_{u}c(x,u),\quad v_n^1(x)=\inf_{u} c(x,u)
\end{align*}
thus we can write,
\begin{align*}
&|v^1(x)-v_n^1(x)|=0.
\end{align*}
For step $k$ we have,
\begin{align*}
&v^k(x)=\inf_{u}\big[c(x,u)+\int_{\mathds{X}}v^{k-1}(x^1)\mathcal{T}(dx^1|x,u)\big]\\
&v_n^k(x)=\inf_{u}\big[c(x,u)+\int_{\mathds{X}}v_n^{k-1}(x^1)\mathcal{T}_n(dx^1|x,u)\big].
\end{align*}
Note that the assumptions of the theorem satisfy the measurable selection criteria and hence we can choose minimizing selectors (\cite[Section 3.3]{HernandezLermaMCP}). If we denote the selectors by $u^*$ and $u_n^*$ we can write
\begin{align*}
&|v^k(x)-v_n^k(x)| \leq \\
& \max\Big(\bigg[|c(x,u^*)-c(x,u^*)|+|\int_{\mathds{X}}v^{k-1}(x^1)\mathcal{T}(dx^1|x,u^*)-\int_{\mathds{X}}v_n^{k-1}(x^1)\mathcal{T}_n(dx^1|x,u^*)|\bigg],\\
& \qquad\bigg[|c(x,u_n^*)-c(x,u_n^*)|+|\int_{\mathds{X}}v^{k-1}(x^1)\mathcal{T}(dx^1|x,u_n^*)-\int_{\mathds{X}}v_n^{k-1}(x^1)\mathcal{T}_n(dx^1|x,u_n^*)|\bigg]\Big)
\end{align*}
For the first term we use \cite[Theorem 20]{royden}. Since $\mathcal{T}_n(\cdot|x,u^*)\to \mathcal{T}(\cdot|x,u^*)$ setwise and $v_n^{k-1}\to v^{k-1}$ pointwise.

For the second term, we use a contradiction argument. Assume that there exists an $\epsilon>0$ and some subsequence (identified with $n_k$) such that 
\begin{align}\label{contr}
|\int_{\mathds{X}}v^{k-1}(x^1)\mathcal{T}(dx^1|x,u^*_{n_k})-\int_{\mathds{X}}v_n^{k-1}(x^1)\mathcal{T}_{n_k}(dx^1|x,u^*_{n_k})|>\epsilon
\end{align}
Now, take a further subsequence $u^*_{n'_k}$ of this sequence which converges to some $u$ whose existence follows from the compactness of $\mathds{U}$. Notice that along this subsequence $\mathcal{T}(\cdot|x,u^*_{n'_k})\to \mathcal{T}(\cdot|x,u)$ and $\mathcal{T}_{n'_k}(\cdot|x,u^*_{n'_k})\to \mathcal{T}(\cdot|x,u)$. Thus, using the induction step and  \cite[Theorem 20]{royden} the above term converges to $0$ along the subsequence indexed by $n'_k$ which contradicts with (\ref{contr}).
The rest of the proof follows from the arguments in Theorem 5.2.
\end{proof}

\subsubsection{Absence of robustness under setwise convergence}
Now, we give a result showing that even if the continuity holds under the setwise convergence of the kernels, the robustness may not be satisfied.
\begin{theorem}\label{notRobustSetwiseFull}
Suppose $\mathcal{T}_n(\cdot|x_n,u_n)\to\mathcal{T}(\cdot|x,u)$ setwise for every $x\in\mathds{X}$ and $u\in\mathds{U}$ and $(x_n,u_n)\to (x,u)$,  then it is not true in general that $J_\beta(\mathcal{T},\gamma_n^*)\to J_\beta(\mathcal{T},\gamma^*)$, even when $\mathds{X}$ and $\mathds{U}$ are compact and $c(x,u)$ is continuous and bounded in $\mathds{X}\times\mathds{U}$.
\end{theorem}
\proof
We prove the result with a counterexample. Define
    \begin{align*}
      L_{n,k}=\left[\frac{2k-2}{2n},\frac{2k-1}{2n}\right),~R_{n,k}=\left[\frac{2k-1}{2n},\frac{k}{n}\right).
    \end{align*}
   Let $L=\left\{y\in\cup_{k=1}^n L_{n,k}\right\}$ and $R=\left\{y\in\cup_{k=1}^n R_{n,k}\right\}$. Next,  define the square-wave function by 
    \begin{align*}
      h_n(t)=1_{\{t\in L\}}-1_{\{t\in R\}},
    \end{align*}
   Define two sequences of probability density functions as
\[f_n(t)=(1+h_n(t))1_{\{t\in [0,1]\}} \quad , \quad g_n(t)=(1-h_n(t))1_{\{t\in [0,1]\}} \]
Consider the kernels given in the following form for $n\geq1$:
\begin{align*}
\mathcal{T}_n(\cdot|x,u)\sim& \mathds{1}_{\{x\in L, u=1\}}f_n(\cdot)+ \mathds{1}_{\{x\in L, u=0\}}g_n(\cdot)\\
&+\mathds{1}_{\{x\in R, u=0\}}f_n(\cdot)+\mathds{1}_{\{x\in R, u=1\}}g_n(\cdot)\\
\mathcal{T}(\cdot|x,u)\sim& U([0,1]).
\end{align*}
By the proof of Riemann-Lebesgue lemma \cite[Theorem 12.21]{wheeden77}
    \begin{align*}
      \lim_{n\to\infty}\int_0^1 h_n(t)g(t) \dd t =0 \text{ for all } g\in L_1\left([0,1],\R\right),
    \end{align*}
   and therefore
    \begin{align*}
      \lim_{n\to\infty}\int_0^1 f_n(t)g(t) \dd t =\int_0^1 g(t) \text{ for all } g\in L_1\left([0,1],\R\right).
    \end{align*}
As a result, $\mathcal{T}_n(\cdot|x,u) \to \mathcal{T}(\cdot|x,u)$ setwise for every $x \in \mathds{X}$ and $u \in \mathds{U}$.

The cost function is given by:
\begin{align*}
c(x,u)=
\begin{cases}
2  &\text{if } u=0\\
x &\text{if } u=1.
\end{cases}
\end{align*}
Notice that if the system starts anywhere on $L$, it does not matter how we define $\gamma_n^*$ for $x\in R$ as the state always stays at $L$.
Thus, with this setup, it can be seen that an optimal policy for $\mathcal{T}_n$ is given by;
\begin{align*}
\gamma_n^*(x)=
\begin{cases}
1  &\text{if } x\in L\\
0  &\text{if } x\in R.
\end{cases}
\end{align*}
when the initial state is $x=0$, which belongs to $L$ for any $n\geq1$. The optimal policy for $\mathcal{T}$ is given by $\gamma^*(x)=1$. 
The discounted cost values can be calculated as follows:
\begin{align*}
J_\beta(\mathcal{T},\gamma^*)=&\sum_{t=0}^{\infty}\beta^t E_{\mathcal{T}}[c(X,1)]=\sum_{t=0}^{\infty}\beta^t \int_0^1xdx=\frac{1}{2(1-\beta)}.
\end{align*}
Building on the calculations in \cite{Prior2017}, the cost under the policy $\gamma_n^*$ is calculated as:
\begin{align*}
J_\beta(\mathcal{T},\gamma_n^*)=&\sum_{t=0}^{\infty}\beta^t E_{\mathcal{T}}[c(X,\gamma^*_n(X))]=\frac{1}{1-\beta}\bigg( \int_L c(x,1)dx+ \int_Rc(x,0)dx\bigg)\\
&=\frac{1}{1-\beta}\bigg(\int_L xdx+\int_R2dx\bigg)=\frac{1}{1-\beta}\bigg(\frac{1}{4}-\frac{1}{8n}+1\bigg) \to \frac{5}{4(1-\beta)}
\end{align*}
which completes the proof. 
\endproof

\subsection{A sufficient condition for robustness under setwise convergence} We now present a similar result to Theorem \ref{robust_fully} that is we show that under the conditions of Theorem \ref{fully_set_suff_cont}, if further for every $n$, $\gamma_n^*$ is optimal for every $x_0\in\mathds{X}$ (under the model $\mathcal{T}_n$) then robustness holds under setwise convergence.

\begin{theorem}\label{robust_fully_set}
Suppose Assumption \ref{set_assmp} holds, if further we have that for every $n$, $\gamma_n^*$ is optimal for every $x_0\in\mathds{X}$ (under the model $\mathcal{T}_n$) then $J_\beta(\mathcal{T},\gamma_n^*)\to J_\beta(\mathcal{T},\gamma^*)$.
\end{theorem}

\begin{proof}
We use the same proof technique as we used for Theorem \ref{robust_fully}. 
Define the following operator for $\gamma_n^*$, an optimal policy for $\mathcal{T}_n$ as in the proof of Theorem \ref{robust_fully},  
\begin{align*}
(T_n(v))(x_0) =   c(x_0,\gamma_n^*(x_0)) + \beta \int v(x_1) \mathcal{T}(dx_1| x_0, \gamma_n^*(x_0))
\end{align*}
which is a contraction from $C_b(\sX)$ to itself under the supremum norm with modulus $\beta$ and has a unique fixed point which is $J_\beta(\mathcal{T},\gamma_n^*)$. Hence, $T_n^k(J_\beta^*(\mathcal{T}))(x)\to J_\beta(\mathcal{T},\gamma_n^*,x)$ as $k\to\infty$ uniformly over $n$, where $T_n^k$ is the operator $T_n$ applied $k$ consecutive times.

Using the properties of setwise convergence we show in Section \ref{tech_operator_conv_set} that

\begin{align}\label{robust_oper_set}
\lim_{n\to\infty}T_n^k(J_\beta^*(\mathcal{T}))(x)= J_\beta^*(\mathcal{T},x),\quad \forall k<\infty.
\end{align}

Then, we write
\begin{align*}
J_\beta(\mathcal{T},\gamma_n^*)-J_\beta^*(\mathcal{T})\leq |J_\beta(\mathcal{T},\gamma_n^*)-T_n^k(J_\beta^*(\mathcal{T}))|+|T_n^k(J_\beta^*(\mathcal{T}))-J_\beta^*(\mathcal{T})|\to 0.
\end{align*}
We can make the second term arbitrarily small by choosing $k$ large enough uniformly over all $n$ since  $T_n^k(J_\beta^*(\mathcal{T}))(x)\to J_\beta(\mathcal{T},\gamma_n^*,x)$ as $k\to\infty$ uniformly over $n$. For the fixed $k$, the first term can be made arbitrarily small by choosing $n$ large enough using (\ref{robust_oper_set}).

\end{proof}


\subsection{Total variation}\label{fully_TV}
The continuity result in Theorem \ref{TV_cont_thm} and the robustness result in Theorem \ref{TV_robust_kernel} apply to this case since the fully observed model may be viewed as a partially observed model with the measurement channel $Q$ given in (\ref{QMDPasPOMDP}).

\begin{remark}
 We note that if the action and state spaces are finite, then total variation convergence and weak convergence coincide and thus Theorem \ref{TV_cont_thm}, Theorem \ref{TV_bound_kernel} and Theorem \ref{TV_robust_kernel} from the partially observed case directly apply to this case considering the channel as a perfect channel. Thus, the only assumptions needed to establish continuity and robustness are 
\begin{itemize}
\item
 $\mathcal{T}_n(\cdot|x,u_n) \to \mathcal{T}(\cdot|x,u)$ in total variation for all $x \in \mathds{X}$. 
\item
$\mathcal{T}(\cdot|x,u)$ is continuous in total variation in $u$ for every given $x \in \mathds{X}$. 
\item
$\mathds{U}$ is compact.
\end{itemize}
Since the spaces are finite, these set of assumptions reduces to 
\begin{itemize}
\item  $\mathcal{T}_n(\cdot|x,u) \to \mathcal{T}(\cdot|x,u)$ in total variation for all $x \in \mathds{X}$ and $u \in \mathds{U}$. 
\end{itemize}

\end{remark}

\begin{remark}
We note that all of the results we present in this paper also apply to finite horizon problems. If we define a finite horizon problem by
\begin{align*}
J(P,\mathcal{T},\gamma)=&\sum_{t=0}^{T} E_{P,Q}^{\mathcal{T}}[c(X_t,U_t)]
\end{align*}
the continuity and robustness properties hold under the same conditions we have presented for the infinite horizon discounted problem.
\end{remark}

\section{Implications for Data-Driven Learning Methods in Stochastic Control}\label{empiricalLearning}

In practice, one might try to learn the kernel of a controlled Markov chain from empirical data; see e.g. \cite{billingsley1961statistical}\cite{Rustem2012}\cite{GyKo07} for some related literature in the control-free and controlled contexts.


Let us briefly discuss the case where a random variable is repeatedly observed, but its probability measure is not known apriori. Let $\{(X_i),\,  i
\in \N \}$ be an $\mathds{X}$-valued i.i.d random variable sequence generated according to some distribution $\mu$.  Defining for every (fixed) Borel $B \subset \mathds{X}$, and $n \in
\mathds{N}$, the empirical occupation measures
\[
\mu_n(B)=
\frac{1}{n}\sum_{i=1}^{n} 1_{\{X_i \in B\}},
\]
one has $\mu_n(B) \to \mu(B)$ almost surely (a.s$.$) by the strong law
of large numbers. Also, $\mu_n \to \mu$ weakly with probability one (\cite{Dudley02}, Theorem 11.4.1). However, $\mu_n$ can not converge to $\mu$ in total variation or setwise, in general. On the other hand, if we know that $\mu$ admits a density, we can find estimators to estimate $\mu$ under total variation \cite[Chapter 3]{Devroye85}. For a more detailed discussion on convergence of empirical occupation measures see \cite[p. 1950-1951]{Prior2017}. In the previous sections, we established robustness results under the convergence of transition kernels in the topology of weak convergence and total variation. We build on these observations next.

\subsection{Application of robustness results to data-driven learning}
\hfill

\begin{corollary}[to Theorem \ref{TV_bound_kernel} and Theorem \ref{TV_robust_kernel} ]
Suppose we are given the following dynamics for finite state space, $\mathds{X}$, and finite action space, $\mathds{U}$,
\[ x_{t+1}= f(x_t,u_t,w_t), \quad  \quad y_t= g(x_t,v_t)\]
where $\{w_t\}$ and $\{v_t\}$ are i.i.d.noise processes and the noise models are unknown.
 Suppose that there is an initial training period so that under some policy, every $x,u$ pair is visited infinitely often if training were to continue indefinitely, but that the training ends at some finite time. Let us assume that, through this training, we empirically learn the transition dynamics with the measurements such that for every (fixed) Borel $B \subset \mathds{X}$, for every $x\in \mathds{X}$, $u \in \mathds{U}$ and $n \in\mathds{N}$, the empirical occupation measures are
\[
\mathcal{T}_n(B|x_0=x,u_0=u)=
\frac{\sum_{i=1}^{n} 1_{\{X_i \in B,X_{i-1}=x,U_{i-1} = u\}}}{\sum_{i=1}^{n} 1_{\{X_{i-1}=x,U_{i-1} = u\}}}.
\]
Then we have that $J_\beta^*(\mathcal{T}_n) \to J_\beta^*(\mathcal{T})$ and $J_\beta(\mathcal{T},\gamma_n^*)\to J_\beta^*(\mathcal{T})$, 
where $\gamma_n^*$ is the optimal policy designed for $\mathcal{T}_n$. Since the channel model $g$ has no restrictions, this result also applies to the fully observed model setup by taking $g(x_t,v_t)=x_t$.
\end{corollary}
\proof
We have that $\mathcal{T}_n(\cdot|x,u) \to \mathcal{T}(\cdot|x,u)$ weakly for every $x \in \mathds{X}$, $u \in \mathds{U}$ almost surely by law of large numbers. Since the spaces are finite, we also have $\mathcal{T}_n(\cdot|x,u) \to \mathcal{T}(\cdot|x,u)$ under total variation. By Theorem \ref{TV_bound_kernel} and Theorem \ref{TV_robust_kernel}, the results follow. \endproof

The following holds for more general spaces.

\begin{corollary}[to  Theorem \ref{weak_robust_kernel}, Theorem \ref{weak_cont}, Theorem \ref{fully_weak_suff_cont} and Theorem \ref{robust_fully} ]\label{emp_cor1}
Suppose we are given the following dynamics with state space $\mathds{X}$ and action space $\mathds{U}$,
\[ x_{t+1}= f(x_t,u_t,w_t), \quad \quad  y_t= g(x_t,v_t)\]
where $\{w_t\}$ and $\{v_t\}$ are i.i.d.noise processes and the noise models are unknown.
Suppose that $f(x,u,\cdot): \mathds{W}\to \mathds{X}$ is invertible for all fixed $(x,u)$ and $f(x,u,w)$ is continuous and bounded on $\mathds{X}\times\mathds{U}\times\mathds{W}$. We construct the empirical measures for the noise process $w_t$  such that for every (fixed) Borel $B \subset \mathds{W}$, and for every $n \in\mathds{N}$, the empirical occupation measures are
\begin{align}\label{inverse_emp}
\mu_n(B)=\frac{1}{n-1}\sum_{i=1}^{n} \mathds{1}_{\{f^{-1}_{x_{i-1},u_{i-1}}(x_i) \in B\} }
\end{align}
where $f^{-1}_{x_{i-1},u_{i-1}}(x_i)$ denotes the inverse of $f(x_{i-1},u_{i-1},w): \mathds{W}\to \mathds{X}$ for given $(x_{i-1},u_{i-1})$. 
Using the noise measurements, we construct the empirical transition kernel estimates for any $(x_0,u_0)$ and Borel $B$ as
\begin{align*}
\mathcal{T}_n(B|x_0,u_0)=\mu_n(f^{-1}_{x_0,u_0}(B)).
\end{align*}
\begin{itemize}
\item [(i)]
  If the measurement channel (represented by the function $g$) is continuous in total variation  then $J_\beta^*(\mathcal{T}_n) \to J_\beta^*(\mathcal{T})$ and $J_\beta(\mathcal{T},\gamma_n^*)\to J_\beta^*(\mathcal{T})$, where $\gamma_n^*$ is the optimal policy designed for $\mathcal{T}_n$ for all initial points.
\item[(ii)]
 If the measurement channel is in the form $g(x_t,v_t)=x_t$ (i.e. fully observed)  then $J_\beta^*(\mathcal{T}_n) \to J_\beta^*(\mathcal{T})$ and if further for every $n$, $\gamma_n^*$ is optimal for every $x_0\in\mathds{X}$ (under the model $\mathcal{T}_n$)  then $J_\beta(\mathcal{T},\gamma_n^*)\to J_\beta^*(\mathcal{T})$.
\end{itemize}
\end{corollary}

\proof
We have $\mu_n \to \mu$ weakly with probability one where $\mu$ is the model. We claim that the transition kernels are such that $\mathcal{T}_n(\cdot|x_n,u_n)\to\mathcal{T}(\cdot|x,u)$ weakly for any $(x_n,u_n) \to (x,u)$. To see that observe the following for $h \in C_b(\mathds{X})$,
\begin{align*}
&\int h(x_1)\mathcal{T}_n(dx_1|x_n,u_n) - \int h(x_1)\mathcal{T}(dx_1|x,u)\\
&=\int h(f(x_n,u_n,w))\mu_n(dw) - \int h(f(x,u,w))\mu(dw)\to0.
\end{align*}
where $\mu_n$ is the empirical measure for $w_t$ and $\mu$ is the true measure again. For the last step, we used that $\mu_n \to \mu$ weakly and $h(f(x_n,u_n,w))$ continuously converge to $h(f(x,u,w))$ i.e. $h(f(x_n,u_n,w_n)) \to h(f(x,u,w)$ for some $w_n \to w$ since $f$ and $g$ are continuous functions. Similarly, it can be also shown that $\mathcal{T}_n(\cdot|x,u)$ and $\mathcal{T}(\cdot|x,u)$ are weakly continuous on $(x,u)$. Thus, for the case where the channel is continuous in total variation by Theorem \ref{weak_robust_kernel} and Theorem \ref{weak_cont} if $c(x,u)$ is bounded and $\mathds{U}$ is compact the result follows.

For the fully observed case,  $J_\beta^*(\mathcal{T}_n) \to J_\beta^*(\mathcal{T})$ by Theorem \ref{fully_weak_suff_cont} and $J_\beta(\mathcal{T},\gamma_n^*)\to J_\beta^*(\mathcal{T})$ by Theorem \ref{robust_fully}.
\endproof

\begin{remark}
We note here that the moment estimation method can also lead to consistency. Suppose that the distribution of $W$ is determined by its moments, such that estimate models $W_n$ have moments of all orders and $\lim_n= E[W_n^r] =E[W^r]$ for all $r\in \Zplus$.
Then, we have that \cite[Thm 30.2]{BillingsleyProbMeasure} $W_n\to W$ weakly and thus $\mathcal{T}_n(\cdot|x_n,u_n)\to\mathcal{T}(\cdot|x,u)$ weakly for any $(x_n,u_n) \to (x,u)$ under the assumptions of above corollary. Hence, we reach continuity and robustness using the same arguments as in the previous result (Corollary \ref{emp_cor1}).
\end{remark}

Now, we give a similar result with the assumption that the noise process of the dynamics admits a continuous probability density function.
\begin{corollary}[to Theorem \ref{TV_bound_kernel} and Theorem \ref{TV_robust_kernel}]\label{emp_cor2}
Suppose we are given the following dynamics for real vector state space $\mathds{X}$ and action space $\mathds{U}$,
\[x_{t+1}= f(x_t,u_t,w_t), \quad \quad y_t= g(x_t,v_t), \]
where $\{w_t\}$ and $\{v_t\}$ are i.i.d.noise processes and the noise models are unknown however it is known that the noise $w_t$ admits a continuous probability density function. Suppose that $f(x,u,\cdot): \mathds{W}\to \mathds{X}$ is invertible for all $(x,u)$. We collect i.i.d. samples of $\{w_t\}$ as in (\ref{inverse_emp}) and use them to construct an estimator, $\tilde{\mu}_n$ , as described in \cite{Devroye85} which consistently estimates $\mu$ in total variation. Using these empirical estimates, we construct the empirical transition kernel estimates for any $(x_0,u_0)$ and Borel $B$ as
\begin{align*}
\mathcal{T}_n(B|x_0,u_0)=\tilde{\mu}_n(f^{-1}_{x_0,u_0}(B)).
\end{align*}
Then independent of the channel, $J_\beta^*(\mathcal{T}_n) \to J_\beta^*(\mathcal{T})$ and $J_\beta(\mathcal{T},\gamma_n^*)\to J_\beta^*(\mathcal{T})$,
where $\gamma_n^*$ is the optimal policy designed for $\mathcal{T}_n$. Since the channel model $g$ has no restrictions, this result also applies to the fully observed model setup by taking $g(x_t,v_t)=x_t$.
\end{corollary}
\proof
By \cite{Devroye85} we can estimate $\mu$ in total variation so that almost surely $\lim_{n \to \infty}\|\tilde{\mu}_n-\mu\|_{TV}=0$. We claim that the convergence of $\tilde{\mu}_n$ to $\mu$ under total variation metric implies the convergence of $\mathcal{T}_n$ to $\mathcal{T}$ in total variation uniformly over all $x \in \mathds{X}$ and $u \in \mathds{U}$ i.e. $\lim_{n\to \infty}\sup_{x,u}\|\mathcal{T}_n(\cdot|x,u)-\mathcal{T}(\cdot|x,u)\|_{TV}=0$. Observe the following,
\begin{align*}
&\sup_{x,u}\|\mathcal{T}_n(\cdot|x,u)-\mathcal{T}(\cdot|x,u)\|_{TV} =\sup_{x,u}\sup_{||h||_\infty\leq1}\big|\int h(x_1)\mathcal{T}_n(dx_1|x,u)-\int h(x_1)\mathcal{T}(dx_1|x,u)\big|\\
&=\sup_{x,u}\sup_{||h||_\infty\leq1}\big|\int h(f(x,u,w))\tilde{\mu}_n(dw) -\int h(f(x,u,w))\mu(dw)\big| \leq \| \tilde{\mu}_n-\mu\|_{TV} \to 0.
\end{align*}
Thus, by Theorem \ref{TV_bound_kernel} and Theorem \ref{TV_robust_kernel}, the result follows.
\endproof

The following example presents some system and channel models which satisfy the requirements of the above corollaries.

\begin{example}
Let $\mathds{X}, \mathds{Y}$, $\mathds{U}$ be real vector spaces with
\[x_{t+1}=f(x_t,u_t)+w_t, \quad \quad y_t= h(x_t,v_t),\]
for unknown i.i.d. noise processes $\{w_t\}$ and $\{v_t\}$.
\begin{enumerate}[label=(\roman*)]
\item
Suppose the channel is in the following form; $y_t= h(x_t,v_t)= x_t + v_t$ where $v_t$ admits a density (e.g. Gaussian density). It can be shown by an application of Scheff\'e's theorem that the channels in this form are continuous in total variation. If further $f(x_t,u_t)$ is continuous and bounded then the requirements of Corollary \ref{emp_cor1} holds for partially observed models.
\item
If the channel is in the following form; $x_t= h(x_t,v_t)$ where $v_t$ admits a density (e.g. Gaussian density) then the system is fully observed. If further $f(x_t,u_t)$ is continuous and bounded then the requirements of Corollary \ref{emp_cor1} holds for fully observed models.
\item
Suppose the function $f(x_t,u_t)$ is known, if the noise process $w_t$ admits a continuous density, then one can estimate the noise model in total variation in a consistent way  (see \cite{Devroye85}). Hence,  the conditions of Corollary \ref{emp_cor2} holds independent of the channel model.
\end{enumerate}
\hfill $\diamond$
\end{example}

\section{Conclusion}
We studied regularity properties of optimal stochastic control on the space of transition kernels, and applications to robustness of optimal control policies designed for an incorrect model applied to an actual system. 


\appendix

\section{Technical Results}
 \subsection{Proof of (\ref{time_t_to_0}) in Theorem \ref{same_policy_cont}}\label{app1}

Before the proof we give a key lemma.
The lemma we present generalizes the following result from \cite[Theorem 3.5]{Langen81} and \cite[Theorem 3.5]{serfozo82}. 

\begin{lemma}\label{langen}
Suppose $\{\mu_n\}_n \subset \P(\mathds{X})$, where $\mathds{X}$ is metric space, converges weakly to some $\mu \in \P(\mathds{X})$. For a bounded real valued sequence of functions $\{f_n\}_n$ such that $\|f_n\|_\infty <C$ for all $n>0$ with $C<\infty$, if $\lim_{n \to \infty}f_n(x_n)=f(x)$ for all $x_n \to x$, i.e. $f_n$ continuously converges to $f$, then
$\lim_{n \to \infty}\int_{\mathds{X}}f_n(x)\mu_n(dx)=\int_{\mathds{X}}f(x)\mu(dx)$.
\end{lemma}

\begin{lemma}\label{kernel}
Suppose we have a uniformly bounded family of functions $\{f_n^\gamma : \mathds{X} \to \R, \gamma\in\Gamma, n>0\}$ such that $\|f_n^\gamma\|_\infty<C$ for all $\gamma \in \Gamma$ and for all $n>0$ for some $C<\infty$.

Further suppose we have another uniformly bounded family of functions $\{f^\gamma : \mathds{X} \to \R, \gamma\in\Gamma\}$ such that $\|f^\gamma\|_\infty<C$ for all $\gamma \in \Gamma$ for some $C<\infty$. Under the following assumptions,
\begin{itemize}
\item[(i)] 
For any $x_n \to x$
\begin{align}
&\sup_{\gamma \in \Gamma}\big|f_n^\gamma(x_n)-f^\gamma(x)\big|\to 0 \label{f1}\\
&\sup_{\gamma \in \Gamma}\big|f^\gamma(x_n)-f^\gamma(x)\big|\to 0. \label{f2}
\end{align}

\item[(ii)]
 $\mathcal{T}_n(\cdot|x_n,u_n)$ converges weakly to $\mathcal{T}(\cdot|x,u)$ for any $(x_n,u_n)\to (x,u)$.
\item[(iii)]
$\mathcal{T}(\cdot|x,u)$ is weakly continuous in $(x,u)$.
\item[(iv)] 
$\mathds{U}$ is compact.
\end{itemize}

 Then for fixed observation realizations, $y_{[0,t]}:= \{y_0,\dots, y_t\}$ and for some $x_t^n \to x_t$
\begin{align}\label{f3}
&\sup_{\gamma \in \Gamma} \bigg|\int \mathcal{T}_n(dx_{t+1}|x_t^n,\gamma(y_{[0,t]}))f_n^\gamma(x_{t+1}) - \int \mathcal{T}(dx_{t+1}|x_t,\gamma(y_{[0,t]}))f^\gamma(x_{t+1})\bigg| \to 0
\end{align}
\end{lemma}
\proof
 Using (\ref{f2}), we see that $\{f^\gamma\}$ is a equicontinuous family of functions. Thus, by the Arzela-Ascoli Theorem, for any given compact set $K \subset \mathds{X}$, and $\epsilon>0$ there is a finite set of continuous functions $\mathds{F}:=\{f^1,\dots,f^N\}$ so that for any $\gamma$, there is $f^i \in \mathds{F}$ with 
\[\sup_{x \in K}|f^\gamma(x)-f^i(x) |\leq \epsilon.\] 
Now, we claim that for the same $\epsilon>0$, the same $f^i \in \mathds{F}$ and the chosen compact set $K$, we can make $\sup_{x\in K} |f_n^\gamma(x)-f^i(x)|\leq 3\epsilon/2$ for large enough $n$ and for any $\gamma \in \Gamma$. To see this, observe the following:
\begin{align*}
&\sup_{x\in K} |f_n^\gamma(x)-f^i(x)|\leq \sup_{x\in K} |f_n^\gamma(x)-f^\gamma(x)|+\sup_{x\in K} |f^\gamma(x)-f^i(x)|
\end{align*}
the second term is less than $\epsilon$ with the argument we made in the first paragraph of the proof. The first term can also be made arbitrarily small since $f_n^\gamma \to f^\gamma$ uniformly on compact sets by (\ref{f1}). Now we wish to show that we can find a compact subset of $\mathds{X}$ such that all probability measures (kernels) in the term (\ref{f3}) put their measure mainly on this compact set. Consider the set of measures $S:=\cup_{\gamma \in \Gamma} S_\gamma$,
where 
\begin{align*}
S_\gamma=\{\mathcal{T}_n(\cdot|x_t^n,\gamma(y_{[0,t]})): \mathcal{T}_n(\cdot|x_t^n,\gamma(y_{[0,t]}))\to \mathcal{T}(\cdot|x_t,\gamma(y_{[0,k]}))\}.
\end{align*}
Here, notice that the set $S$ depends on the sequence $\{x_t^n\}$ and the observation realizations $y_{[0,t]}$. To cover all the kernels in (\ref{f3}) we take $\{x_t^n\}$ and $y_{[0,t]}$ as they are given in the statement of the lemma. 

For a sequence from the set $S$, say $\mathcal{T}_{n_m}(\cdot|x_t^{n_m},\gamma_m(y_{[0,t]}))$; since $\mathds{U}$ is a compact set and the observations are fixed, there exists a subsequence in which $\gamma_{m_r}(y_{[0,t]}) \to u^*$ for some $u^* \in \mathds{U}$. Now we focus on this subsequence $\mathcal{T}_{n_{m_r}}(\cdot|x_t^{n_{m_r}},\gamma_{m_r}(y_{[0,t]}))$.
By the assumption $(ii)$ in the lemma statement $\{\mathcal{T}_n\}_n$ also satisfies the following: for any $(x_n,u_n)\to (x,u)$, $\mathcal{T}_n(\cdot|x_n,u_n) \to \mathcal{T}(\cdot|x,u)$. Thus,
\begin{align*}
\mathcal{T}_{n_{m_r}}(\cdot|x_t^{n_{m_r}},\gamma_{m_r}(y_{[0,t]})) \to \mathcal{T}(\cdot|x_t,u^*).
\end{align*}
Hence any sequence in the set $S$ has a convergent subsequence, thus $S$ is a precompact family. Therefore, it is a tight family of functions by Prokhorov theorem \cite[Theorem 5.2]{Billingsley} (see also \cite[Theorem 11.5.3]{Dudley02}). Hence, for any $\epsilon >0$, there exists a compact set $K_\epsilon$ such that for all $n$ and uniformly for all $\gamma \in \Gamma$, 
\[\int_{K_\epsilon} \mathcal{T}_n(dx_1|x_t^n,\gamma(y_{[0,t]})) \geq 1-\epsilon.\]


Now, we fix an $\epsilon>0$, choose a compact set $K_\epsilon$ according to above discussion such that all $\mathcal{T}_n$ puts almost all their measure (more than $1-\epsilon$)  on $K_\epsilon$.We also fix a finite family of continuous functions $\mathds{F}:=\{f^1,\dots,f^{N}\}$ such that for any $\gamma$, we can find an $f^i \in \mathds{F}$ with $\sup_{x_t \in K_\epsilon}|f^\gamma(x_t)-f^i(x_t) |\leq \epsilon$. Moreover, we choose a large $N$ such that $\sup_{x\in K_\epsilon} |f_n^\gamma(x)-f^i(x)|\leq 3\epsilon/2$ for all $n\geq N$.

With this setup, we go back to the main statement:
\begin{align*}
&\sup_{\gamma \in \Gamma} \bigg|\int \mathcal{T}_n(dx_{t+1}|x_t^n,\gamma(y_{[0,t]}))f_n^\gamma(x_{t+1}) - \int \mathcal{T}(dx_{t+1}|x_t,\gamma(y_{[0,t]}))f^\gamma(x_{t+1})\bigg|\\
&\leq\sup_{\gamma \in \Gamma}\big|\int_{\mathds{X}\setminus K_\epsilon} \mathcal{T}_n(dx_{t+1}|x_t^n,\gamma(y_{[0,t]}))f_n^\gamma(x_{t+1})- \int_{\mathds{X}\setminus K_\epsilon} \mathcal{T}(dx_{t+1}|x_t,\gamma(y_{[0,t]}))f^\gamma(x_{t+1})\big|\\
&\qquad+\sup_{\gamma \in \Gamma}\big|\int_{ K_\epsilon} \mathcal{T}_n(dx_{t+1}|x_t^n,\gamma(y_{[0,t]}))f_n^\gamma(x_{t+1})- \int_{ K_\epsilon} \mathcal{T}(dx_{t+1}|x_t,\gamma(y_{[0,t]}))f^\gamma(x_{t+1})\big|\\
&\leq 2\epsilon C + \sup_{\gamma \in \Gamma}\big|\int_{K_\epsilon} \mathcal{T}_n(dx_{t+1}|x_{t}^n,\gamma(y_{[0,t]}))\big(f_n^\gamma(x_{t+1}) -f^i(x_{t+1})\big)\\
&\qquad\qquad\qquad + \int_{K_\epsilon} \mathcal{T}_n(dx_{t+1}|x_{t}^n,\gamma(y_{[0,t]}))f^i(x_{t+1}) - \int_{K_\epsilon} \mathcal{T}(dx_{t+1}|x_t,\gamma(y_{[0,t]}))f^i(x_{t+1})\\
&\qquad\qquad\qquad\qquad + \int_{K_\epsilon} \mathcal{T}(dx_{t+1}|x_t,\gamma(y_{[0,t]}))\big(f^i(x_{t+1})-f^\gamma(x_{t+1})\big)\big|\\
&\leq 2\epsilon C + \sup_{\gamma \in \Gamma} \big| \int_{K_\epsilon} \mathcal{T}_n(dx_{t+1}|x_{t}^n,\gamma(y_{[0,t]}))f^i(x_{t+1})\\
&\qquad\qquad\qquad\qquad - \int_{K_\epsilon} \mathcal{T}(dx_{t+1}|x_t,\gamma(y_{[0,t]}))f^i(x_{t+1})\big| + 5\epsilon/2 \leq 4\epsilon C + 7\epsilon/2
\end{align*}
where $C$ is the uniform bound of $f_n^\gamma$ and $f^i(x_{t+1})$ is chosen according to the discussion above such that $f^i$ is $\epsilon$ close to $f^\gamma(x_{t+1})$ and the same $f^i$ is $3\epsilon/2$ close to $f_n^\gamma(x_{t+1})$. 

At the last step, we used the fact that $\mathcal{T}_n(dx_{t+1}|x_{t}^n,\gamma(y_{[0,t]}))$ converges weakly to $\mathcal{T}(dx_{t+1}|x_{t},\gamma(y_{[0,t]}))$ uniformly on $\mathds{U}$. Thus,
\begin{align*}
&\sup_{\gamma \in \Gamma} \big| \int_{K_\epsilon} \mathcal{T}_n(dx_{t+1}|x_{t}^n,\gamma(y_{[0,t]}))f^i(x_{t+1}) - \int_{K_\epsilon} \mathcal{T}(dx_{t+1}|x_t,\gamma(y_{[0,t]}))f^i(x_{t+1})\big|\\
&\leq\sup_{\gamma \in \Gamma} \big| \int_{\mathds{X}- K_\epsilon} \mathcal{T}_n(dx_{t+1}|x_{t}^n,\gamma(y_{[0,t]}))f^i(x_{t+1}) - \int_{\mathds{X}-K_\epsilon} \mathcal{T}(dx_{t+1}|x_t,\gamma(y_{[0,t]}))f^i(x_{t+1})\big|\\
&+\sup_{\gamma \in \Gamma} \big| \int_{\mathds{X}} \mathcal{T}_n(dx_{t+1}|x_{t}^n,\gamma(y_{[0,t]}))f^i(x_{t+1}) - \int_{\mathds{X}} \mathcal{T}(dx_{t+1}|x_t,\gamma(y_{[0,t]}))f^i(x_{t+1})\big| \leq 2\epsilon C + \epsilon
\end{align*}
for large enough $n$. As $\epsilon$ is arbitrary, the result follows.
\endproof

With this lemma, we go back to (\ref{time_t_to_0}).
For easiness of notation we will first study the case where $t=3$.
\begin{align*}
&\sup_{\gamma\in \Gamma}\Big|E_P^{{\mathcal{T}}}\bigg[c\big(X_3,\gamma(Y_{[0,3]})\big)\bigg]-E_P^{{\mathcal{T}_n}}\bigg[c\big(X_3,\gamma(Y_{[0,3]})\big)\bigg]\Big|\\
&=\sup_{\gamma\in \Gamma}\big|\int P(dx_0)Q(dy_0|x_0)\mathcal{T}(dx_1|x_0,\gamma(y_0))Q(dy_1|x_1)\mathcal{T}(dx_2|x_1,\gamma(y_{[0,1]}))\\
&\qquad\qquad \times Q(dy_2|x_2)\mathcal{T}(dx_3|x_2,\gamma(y_{[0,2]}))Q(dy_3|x_3)c(x_3,\gamma(y_{[0,3]}))\\
&\quad-\int P(dx_0)Q(dy_0|x_0)\mathcal{T}_n(dx_1|x_0,\gamma(y_0))Q(dy_1|x_1)\mathcal{T}_n(dx_2|x_1,\gamma(y_{[0,1]}))\\
&\qquad\qquad \times Q(dy_2|x_2)\mathcal{T}_n(dx_3|x_2,\gamma(y_{[0,2]}))Q(dy_3|x_3)c(x_3,\gamma(y_{[0,3]}))\big|.
\end{align*}
Using the dominated convergence theorem, it suffices to show that
\begin{align*}
&\sup_{\gamma\in \Gamma} \bigg|\int \mathcal{T}_n(dx_1|x_0,\gamma(y_0))Q(dy_1|x_1)\mathcal{T}_n(dx_2|x_1,\gamma(y_{[0,1]}))\\
&\qquad\qquad \times Q(dy_2|x_2)\mathcal{T}_n(dx_3|x_2,\gamma(y_{[0,2]}))Q(dy_3|x_3)c(x_3,\gamma(y_{[0,3]}))\\
&\qquad - \int \mathcal{T}(dx_1|x_0,\gamma(y_0))Q(dy_1|x_1)\mathcal{T}(dx_2|x_1,\gamma(y_{[0,1]}))\\
&\qquad\qquad \times Q(dy_2|x_2)\mathcal{T}(dx_3|x_2,\gamma(y_{[0,2]}))Q(dy_3|x_3)c(x_3,\gamma(y_{[0,3]}))\bigg| \to 0.
\end{align*}
Then, using Lemma \ref{kernel}, it suffices to show that for any $x_1^n \to x_1$
\begin{align}\label{vary_k}
\sup_{\gamma\in \Gamma} \bigg|\int Q(dy_1|x_1^n)\mathcal{T}_n(dx_2|x_1^n,\gamma(y_{[0,1]})) Q(dy_2|x_2)&\mathcal{T}_n(dx_3|x_2,\gamma(y_{[0,2]}))\nonumber\\
&Q(dy_3|x_3)c(x_3,\gamma(y_{[0,3]}))\nonumber\\
 - \int Q(dy_1|x_1)\mathcal{T}(dx_2|x_1,\gamma(y_{[0,1]}))Q(dy_2|x_2)&\mathcal{T}(dx_3|x_2,\gamma(y_{[0,2]}))\nonumber\\
&Q(dy_3|x_3)c(x_3,\gamma(y_{[0,3]}))\bigg| \to 0
\end{align}
and
\begin{align}\label{fix_k}
\sup_{\gamma\in \Gamma} \bigg|\int Q(dy_1|x_1^n)\mathcal{T}(dx_2|x_1^n,\gamma(y_{[0,1]})) Q(dy_2|x_2)&\mathcal{T}(dx_3|x_2,\gamma(y_{[0,2]}))\nonumber\\
&Q(dy_3|x_3)c(x_3,\gamma(y_{[0,3]}))\nonumber\\
- \int Q(dy_1|x_1)\mathcal{T}(dx_2|x_1,\gamma(y_{[0,1]})) Q(dy_2|x_2)&\mathcal{T}(dx_3|x_2,\gamma(y_{[0,2]}))\nonumber\\
&Q(dy_3|x_3)c(x_3,\gamma(y_{[0,3]}))\bigg| \to 0.
\end{align}
We only focus on the term (\ref{vary_k}), the analysis for the term (\ref{fix_k}) follows from identical steps.
For (\ref{vary_k}), we write the following:
\begin{align}\label{lem_step}
\sup_{\gamma\in \Gamma} \bigg|\int Q(dy_1|x_1^n)\mathcal{T}_n(dx_2|x_1^n,\gamma(y_{[0,1]})) Q(dy_2|x_2)&\mathcal{T}_n(dx_3|x_2,\gamma(y_{[0,2]}))\nonumber\\
&Q(dy_3|x_3)c(x_3,\gamma(y_{[0,3]}))\nonumber\\
\qquad - \int Q(dy_1|x_1)\mathcal{T}(dx_2|x_1,\gamma(y_{[0,1]})) Q(dy_2|x_2)&\mathcal{T}(dx_3|x_2,\gamma(y_{[0,2]}))\nonumber\\
&Q(dy_3|x_3)c(x_3,\gamma(y_{[0,3]}))\bigg|\nonumber\\
\leq \sup_{\gamma\in \Gamma} \bigg|\int Q(dy_1|x_1^n)\mathcal{T}_n(dx_2|x_1^n,\gamma(y_{[0,1]})) Q(dy_2|x_2)&\mathcal{T}_n(dx_3|x_2,\gamma(y_{[0,2]}))\nonumber\\
&Q(dy_3|x_3)c(x_3,\gamma(y_{[0,3]}))\nonumber\\
\qquad - \int Q(dy_1|x_1)\mathcal{T}_n(dx_2|x_1^n,\gamma(y_{[0,1]})) Q(dy_2|x_2)&\mathcal{T}_n(dx_3|x_2,\gamma(y_{[0,2]}))\nonumber\\
&Q(dy_3|x_3)c(x_3,\gamma(y_{[0,3]}))\bigg|\nonumber\\
\quad+\sup_{\gamma\in \Gamma} \bigg|\int Q(dy_1|x_1)\mathcal{T}_n(dx_2|x_1^n,\gamma(y_{[0,1]})) Q(dy_2|x_2)&\mathcal{T}_n(dx_3|x_2,\gamma(y_{[0,2]}))\nonumber\\
&Q(dy_3|x_3)c(x_3,\gamma(y_{[0,3]}))\nonumber\\
\qquad - \int Q(dy_1|x_1)\mathcal{T}(dx_2|x_1,\gamma(y_{[0,1]})) Q(dy_2|x_2)&\mathcal{T}(dx_3|x_2,\gamma(y_{[0,2]}))\nonumber\\
&Q(dy_3|x_3)c(x_3,\gamma(y_{[0,3]}))\bigg|\end{align}
The first term goes to 0 since the channel is continuous in total variation. For the second term, using Lemma \ref{kernel} and the total variation continuity of $Q$ successively, it reduces to show that
\begin{align*}
&\sup_{\gamma\in \Gamma} \bigg|\int Q(dy_3|x_3^n)c(x_3^n,\gamma(y_{[0,3]}) - \int Q(dy_3|x_3)c(x_3,\gamma(y_{[0,3]})\bigg|\to 0.
\end{align*}
To show this, we write the following:
\begin{align*}
&\sup_{\gamma\in \Gamma} \bigg|\int Q(dy_3|x_3^n)c(x_3^n,\gamma(y_{[0,3]}) - \int Q(dy_3|x_3)c(x_3,\gamma(y_{[0,3]})\bigg|\\
&\leq  \sup_{\gamma\in \Gamma} \bigg| \int Q(dy_3|x_3^n)c(x_3^n,\gamma(y_{[0,3]}) -  \int Q(dy_3|x_3)c(x_3^n,\gamma(y_{[0,3]})\bigg|\\
&\qquad+ \sup_{\gamma\in \Gamma} \int Q(dy_3|x_3)\big|c(x_3^n,\gamma(y_{[0,3]}) - c(x_3,\gamma(y_{[0,3]})\big|.
\end{align*}
The first term goes to 0 since $Q$ is continuous in total variation and the second term goes to 0 since $c$ is continuous in $x$ uniformly over $\mathds{U}$. Thus, (\ref{vary_k}) holds true. (\ref{fix_k}) also holds true with identical arguments; we use the convergence of $\mathcal{T}(\cdot|x_n,u)$ to $\mathcal{T}(\cdot|x,u)$ uniformly over $u \in\mathds{U}$ whereas for (\ref{vary_k}), we use the convergence of $\mathcal{T}_n(\cdot|x_n,u)$ to $\mathcal{T}(\cdot|x,u)$ uniformly over $u \in\mathds{U}$ at (\ref{lem_step}) with Lemma \ref{kernel}. Therefore, (\ref{time_t_to_0}) goes to 0 for the time step $t=3$. For a general finite time stage $t$, we can again use the iterative approach we used when $t=3$. Thus,
 we can generalize that for any $0<t<\infty$
\begin{align*}
\sup_{\gamma \in \Gamma}\bigg|E_P^{{\mathcal{T}}}\Big[c\big(X_t,\gamma(Y_{[0,t]})\big)\Big]-E_P^{{\mathcal{T}_n}}\Big[c\big(X_t,\gamma(Y_{[0,t]})\big)\Big]\bigg|\to 0.
\end{align*}

 \subsection{Proof of (\ref{TV_time_t_to_0}) in Theorem \ref{TV_cont_thm}}\label{app2}

Before the proof we give a key lemma.
\begin{lemma}\label{TV_vary}
For a uniformly bounded family of functions $\{f_n^\gamma:\mathds{X}\to\R, n>0, \gamma\in\Gamma\}$ and $\{f^\gamma:\mathds{X}\to\R,\gamma\in\Gamma\}$ if we have $\sup_{\gamma \in \Gamma}\big|f_n^\gamma(x)-f^\gamma(x)\big|\to 0$, then
\begin{align*}
&\sup_{\gamma \in \Gamma} \bigg|\int \mathcal{T}_n(dx_{t+1}|x_t,\gamma(y_{[0,t]}))f_n^\gamma(x_{t+1}) - \int \mathcal{T}(dx_{t+1}|x_t,\gamma(y_{[0,t]}))f^\gamma(x_{t+1})\bigg| \to 0
\end{align*}
for a fixed observation realizations $y_{[0,t]}:=\{y_0,\dots,y_t\}$ and a fixed state $x_t$, under the following assumptions
\begin{itemize}
\item[(i)]
 $\mathcal{T}_n$ is such that for any sequence $\{u_n\} \subset \mathds{U}$ converging to some $u \in \mathds{U}$, $\mathcal{T}_n(\cdot|x,u_n) \to \mathcal{T}(\cdot|x,u)$ in total variation for all $x \in \mathds{X}$,
\item[(ii)]
$\mathcal{T}(\cdot|x,u)$ is continuous in total variation in $u$ for every given $x \in \mathds{X}$.
\end{itemize}
\end{lemma}
\proof
\begin{align*}
&\sup_{\gamma \in \Gamma} \bigg|\int \mathcal{T}_n(dx_{t+1}|x_t,\gamma(y_{[0,t]}))f_n^\gamma(x_{t+1}) - \int \mathcal{T}(dx_{t+1}|x_t,\gamma(y_{[0,t]}))f^\gamma(x_{t+1})\bigg| \\
&\leq\sup_{\gamma \in \Gamma} \bigg|\int \mathcal{T}_n(dx_{t+1}|x_t,\gamma(y_{[0,t]}))f_n^\gamma(x_{t+1}) - \int \mathcal{T}(dx_{t+1}|x_t,\gamma(y_{[0,t]}))f_n^\gamma(x_{t+1})\bigg| \\
&\qquad+ \sup_{\gamma \in \Gamma} \bigg|\int \mathcal{T}(dx_{t+1}|x_t,\gamma(y_{[0,t]}))\big(f_n^\gamma(x_{t+1}) -f^\gamma(x_{t+1})\big)\bigg|\\
&\leq \sup_{\gamma \in \Gamma}\| \mathcal{T}_n(dx_{t+1}|x_t,\gamma(y_{[0,t]}))- \mathcal{T}(dx_{t+1}|x_t,\gamma(y_{[0,t]}))\|_{TV}\\
&\qquad+ \sup_{\gamma \in \Gamma} \bigg|\int \mathcal{T}(dx_{t+1}|x_t,\gamma(y_{[0,t]}))\big(f_n^\gamma(x_{t+1}) -f^\gamma(x_{t+1})\big)\bigg|.
\end{align*}
Above, the first term goes to 0 as $\mathcal{T}_n(\cdot|x,u_n) \to \mathcal{T}(\cdot|x,u)$ in total variation and $\mathds{U}$ is compact. 

For the second term, first we use the  assumption that $\mathcal{T}(\cdot|x,u)$ is continuous in $u$. For any $\epsilon>0$, there exists a $\delta>0$ such that $|u'-u|<\delta$ implies $\|\mathcal{T}(\cdot|x,u)-\mathcal{T}(\cdot|x,u')\|_{TV}<\epsilon$. Furthermore, by the assumption $\mathds{U}$ is compact. Therefore, for the given $\delta$, there exists a finite set $\{u_1,\cdots,u_N\}$ such that for any $\gamma \in \Gamma$, we can find a $u_i$ with $|u_i-\gamma(y_{[0,t]})|<\delta$. 

Combining what we have; for any $\epsilon>0$ and for any $\gamma \in \Gamma$, we can find a $u_i$ such that $\|\mathcal{T}(\cdot|x,\gamma(y_{[0,t]}))-\mathcal{T}(\cdot|x,u_i)\|_{TV}<\epsilon$. Now we focus on the second term again:
\begin{align*}
&\sup_{\gamma \in \Gamma} \bigg|\int \mathcal{T}(dx_{t+1}|x_t,\gamma(y_{[0,t]}))\big(f_n^\gamma(x_{t+1}) -f^\gamma(x_{t+1})\big)\bigg|\\
&\leq\sup_{\gamma \in \Gamma} \bigg|\int \mathcal{T}(dx_{t+1}|x_t,\gamma(y_{[0,t]}))\big(f_n^\gamma(x_{t+1}) -f^\gamma(x_{t+1})\big) \\
&\qquad\qquad\qquad\qquad -\int \mathcal{T}(dx_{t+1}|x_t,u_i)\big(f_n^\gamma(x_{t+1}) -f^\gamma(x_{t+1})\big)\bigg|\\
&\qquad+\sup_{\gamma \in \Gamma}\int \mathcal{T}(dx_{t+1}|x_t,u_i)\big|f_n^\gamma(x_{t+1}) -f^\gamma(x_{t+1})\big|\\
&\leq \|c\|_\infty \sup_{\gamma \in \Gamma} \|\mathcal{T}(\cdot|x,\gamma(y_{[0,t]}))-\mathcal{T}(\cdot|x,u_i)\|_{TV}\\
&\qquad\qquad\qquad +\sup_{\gamma \in \Gamma}\int \mathcal{T}(dx_{t+1}|x_t,u_i)\big|f_n^\gamma(x_{t+1}) -f^\gamma(x_{t+1})\big|
\end{align*}
where $\|c\|_\infty$ is a uniform bound of $f_n$ and $u_i$ is chosen according to the above discussion. Thus, the first term is less than $\|c\|_\infty \epsilon$ and the second term can be made arbitrarily small for large enough $n$ since $\sup_{\gamma \in \Gamma}\big|f_n^\gamma(x)-f^\gamma(x)\big|\to 0$ by assumption. The result follows.
\endproof

Now we go back to (\ref{TV_time_t_to_0}). We will first study the case where $t=3$.
\begin{align*}
&\sup_{\gamma \in \Gamma}\bigg|E_P^{{\mathcal{T}}}\Big[c\big(X_3,\gamma(Y_{[0,3]})\big)\Big]-E_P^{{\mathcal{T}_n}}\Big[c\big(X_3,\gamma(Y_{[0,3]})\big)\Big]\bigg|\\
&=\sup_{\gamma \in \Gamma}\bigg|\int P(dx_0)Q(dy_0|x_0)\mathcal{T}(dx_1|x_0,\gamma(y_0))Q(dy_1|x_1)\mathcal{T}(dx_2|x_1,\gamma(y_{[0,1]}))\\
&\qquad\qquad \times Q(dy_2|x_2)\mathcal{T}(dx_3|x_2,\gamma(y_{[0,2]}))Q(dy_3|x_3)c(x_3,\gamma(y_{[0,3]}))\\
&\quad-\int P(dx_0)Q(dy_0|x_0)\mathcal{T}_n(dx_1|x_0,\gamma(y_0))Q(dy_1|x_1)\mathcal{T}_n(dx_2|x_1,\gamma(y_{[0,1]}))\\
&\qquad\qquad \times Q(dy_2|x_2)\mathcal{T}_n(dx_3|x_2,\gamma(y_{[0,2]}))Q(dy_3|x_3)c(x_3,\gamma(y_{[0,3]}))\bigg|.
\end{align*}
Using Lemma \ref{TV_vary} it suffices to show that
\begin{align*}
&\sup_{\gamma \in \Gamma}\bigg|\int \bigg( Q(dy_1|x_1)\mathcal{T}_n(dx_2|x_1,\gamma(y_{[0,1]}))Q(dy_2|x_2)\mathcal{T}_n(dx_3|x_2,\gamma(y_{[0,2]}))\\
&\qquad -  Q(dy_1|x_1)\mathcal{T}(dx_2|x_1,\gamma(y_{[0,1]}))Q(dy_2|x_2)\mathcal{T}(dx_3|x_2,\gamma(y_{[0,2]})) \bigg)\\
&\qquad \qquad \qquad \qquad \qquad \times Q(dy_3|x_3) c(x_3,\gamma(y_{[0,3]}))\bigg| \to 0.
\end{align*}
Following the same procedure and using Lemma \ref{TV_vary} successively, it reduces to show that 
\begin{align*}
&\sup_{\gamma \in \Gamma}\bigg| \int \mathcal{T}_n(dx_3|x_2,\gamma(y_{[0,2]})) Q(dy_3|x_3)c(x_3,\gamma(y_{[0,3]}))\\
&\qquad\qquad - \int \mathcal{T}(dx_3|x_2,\gamma(y_{[0,2]})) Q(dy_3|x_3)c(x_3,\gamma(y_{[0,3]}))\bigg|\\
&\leq \|c\|_\infty\sup_{\gamma \in \Gamma} \| \mathcal{T}_n(dx_3|x_2,\gamma(y_{[0,2]}))-  \mathcal{T}(dx_3|x_2,\gamma(y_{[0,2]}))\|_{TV}\to0
\end{align*}
which holds true by the assumptions, i.e., since the action space $\mathds{U}$ is compact and $\mathcal{T}_n$ is such that for any sequence $\{u_n\} \subset \mathds{U}$ converging to some $u \in \mathds{U}$, $\mathcal{T}_n(\cdot|x,u_n) \to \mathcal{T}(\cdot|x,u)$ in total variation for all $x \in \mathds{X}$. This argument can be applied to any time step $t < \infty$. 

 \subsection{Proof of (\ref{TV_sup_bound}) in Theorem \ref{TV_bound_kernel}}\label{app3}
First, we provide the analysis for $k=2$.
\begin{align*}
&\|P^\gamma_{ \mathcal{T}_n}(d(x,y,u)_{[0,2]})-P^\gamma_{ \mathcal{T}}(d(x,y,u)_{[0,2]})\|_{TV}\\
&=\sup_{||f||_{\infty}\leq 1} \bigg|\int  P(\dd x_0) Q(\dd y_0|x_0) 1_{\{\gamma(y_0) \in \dd u_0\}} \mathcal{T}_n(\dd x_1|x_0,u_0) Q(\dd y_1|x_1) 1_{\{\gamma(y_0,y_1) \in \dd u_1\}}\\
&\qquad\qquad\qquad \times  \mathcal{T}_n(\dd x_2|x_1,u_1) Q(\dd y_2|x_2) 1_{\{\gamma(y_0,y_1,y_2) \in \dd u_2\}} f(x,y,u)_{[0,2]}\\
&\qquad\qquad - \int  P(\dd x_0) Q(\dd y_0|x_0) 1_{\{\gamma(y_0) \in \dd u_0\}} \mathcal{T}(\dd x_1|x_0,u_0) Q(\dd y_1|x_1) 1_{\{\gamma(y_0,y_1) \in \dd u_1\}}\\
&\qquad\qquad\qquad \times  \mathcal{T}(\dd x_2|x_1,u_1) Q(\dd y_2|x_2) 1_{\{\gamma(y_0,y_1,y_2) \in \dd u_2\}} f(x,y,u)_{[0,2]} \bigg|\\
&\leq \sup_{||f||_{\infty}\leq 1}\int P^\gamma_{ \mathcal{T}_n}(d(x_{[0,1]},y_{[0,1]},u_{[0,1]})) \\
&\qquad\quad\quad\bigg|\int \mathcal{T}_n(\dd x_2|x_1,u_1) Q(\dd y_2|x_2) 1_{\{\gamma(y_{[0,2]}) \in \dd u_2\}} f(x,y,u)_{[0,2]}\\
&\qquad\quad\qquad\qquad - \int \mathcal{T}(\dd x_2|x_1,u_1) Q(\dd y_2|x_2) 1_{\{\gamma(y_{[0,2]}) \in \dd u_2\}} f(x,y,u)_{[0,2]} \bigg|\\
&+ \sup_{||f||_{\infty}\leq 1}\int P^\gamma_{ \mathcal{T}}(d(x_0,y_0,u_0))\\
&\qquad\quad \bigg|\int \mathcal{T}_n(\dd x_1|x_0,u_0) Q(\dd y_1|x_1) 1_{\{\gamma(y_0,y_1) \in \dd u_1\}} P^\gamma_{ \mathcal{T}}(d(x_2,y_2,u_2))f(x,y,u)_{[0,2]}\\
&\qquad\quad\quad - \int \mathcal{T}(\dd x_1|x_0,u_0) Q(\dd y_1|x_1) 1_{\{\gamma(y_0,y_1) \in \dd u_1\}} P^\gamma_{ \mathcal{T}}(d(x_2,y_2,u_2))f(x,y,u)_{[0,2]} \bigg|\\
&\leq 2  \sup_{x \in \mathds{X}, u \in \mathds{U}} \| \mathcal{T}_n(.|x,u)- \mathcal{T}(.|x,u)\|_{TV}
\end{align*}
 Now, we do the same analysis for a general time step $k$:
\begin{align*}
&\|P^\gamma_{ \mathcal{T}_n}(d(x,y,u)_{[0,k]})-P^\gamma_{ \mathcal{T}}(d(x,y,u)_{[0,k]})\|_{TV}\\
&=\sup_{||f||_{\infty}\leq 1} \bigg| \int f(x,y,u)_{[0,k]}P^\gamma_{ \mathcal{T}_n}(d(x,y,u)_{[0,k]})-\int f(x,y,u)_{[0,k]}P^\gamma_{ \mathcal{T}}(d(x,y,u)_{[0,k]}) \bigg|\\
&\leq\sup_{||f||_{\infty}\leq 1} \bigg|\int P^\gamma_{ \mathcal{T}, \mathcal{T}_n}(dx_0,dy_0,du_0)\\
&\qquad\qquad\bigg[\int  \mathcal{T}(dx_1|x_0,u_0) \int f(x,y,u)_{[0,k]} P^\gamma_{ \mathcal{T}, \mathcal{T}_n}(dx_{[2,k]},dy_{[1,k]},du_{[1,k]})\\
&\hspace{2.5cm}-\int  \mathcal{T}_n(dx_1|x_0,u_0) \int f(x,y,u)_{[0,k]} P^\gamma_{ \mathcal{T}, \mathcal{T}_n}(dx_{[2,k]},dy_{[1,k]},du_{[1,k]})\bigg] \bigg|\\
&+ \bigg |\int  P^\gamma_{ \mathcal{T}, \mathcal{T}_n}(dx_{[0,1]},dy_{[0,1]},du_{[0,1]})\\
&\qquad\qquad\bigg[\int  \mathcal{T}(dx_2|x_1,u_1) \int f(x,y,u)_{[0,k]} P^\gamma_{ \mathcal{T}, \mathcal{T}_n}(dx_{[3,k]},dy_{[2,k]},du_{[3,k]})\\
&\hspace{2.5cm}-\int  \mathcal{T}_n(dx_2|x_1,u_1) \int f(x,y,u)_{[0,k]} P^\gamma_{ \mathcal{T}, \mathcal{T}_n}(dx_{[3,k]},dy_{[2,k]},du_{[2,k]})\bigg] \bigg|\\
&\dots+ \bigg|\int  P^\gamma_{ \mathcal{T}, \mathcal{T}_n}(dx_{[0,k-1]},dy_{[0,k-1]},du_{[0,k-1]})\bigg[\int  \mathcal{T}(dx_k|x_{k-1},u_{k-1})  f(x,y,u)_{[0,k]} \\
&\hspace{6.4cm}-\int  \mathcal{T}_n(dx_k|x_{k-1},u_{k-1}) f(x,y,u)_{[0,k]}\bigg] \bigg|\\
&\leq \int P^\gamma_{ \mathcal{T}, \mathcal{T}_n}(dx_0,dy_0,du_0)\| \mathcal{T}(\cdot|x_0,u_0)- \mathcal{T}_n(\cdot|x_0,u_0)\|_{TV}\\
&\quad+\int  P^\gamma_{ \mathcal{T}, \mathcal{T}_n}(dx_{[0,1]},dy_{[0,1]},du_{[0,1]})\| \mathcal{T}(\cdot|x_1,u_1)- \mathcal{T}_n(\cdot|x_1,u_1)\|_{TV}\\
&\dots+\int  P^\gamma_{ \mathcal{T}, \mathcal{T}_n}(dx_{[0,k-1]},dy_{[0,k-1]},du_{[0,k-1]}))\| \mathcal{T}(\cdot|x_{k-1},u_{k-1})- \mathcal{T}_n(\cdot|x_{k-1},u_{k-1})\|_{TV}\\
&\leq k  \sup_{x \in \mathds{X}, u \in \mathds{U}} \| \mathcal{T}(.|x,u)- \mathcal{T}_n(.|x,u)\|_{TV}.
\end{align*}
In the argument above $P^\gamma_{ \mathcal{T}, \mathcal{T}_n}$ denotes a strategic measure that uses $\mathcal{T}$ or $\mathcal{T}_n$ at various steps. The terms are arranged so that at every term the applied strategic measures coincide. 

\subsection{Proof for Remark \ref{equ_remark}}\label{equ_fully}
We note that if the family where we search for policies is restricted to an equicontinuous family of functions, robustness can also be achieved. Let $\Gamma_{eq}$ be the family of equicontinuous policies so that for any given $x_0 \in \mathds{X}$ and $\epsilon>0$, there exists a $\delta>0$ such that $|\gamma(x) - \gamma(x_0)| \leq \epsilon$ for all $\gamma\in \Gamma_{eq}$ and for every $x$ such that $|x-x_0| \leq \delta$. 

 We show that for all $t<\infty$
\begin{align*}
&\sup_{\gamma \in \Gamma_{eq}}\bigg|E^{{\mathcal{T}}}\Big[c\big(X_t,\gamma(X_{t})\big)\Big]-E^{{\mathcal{T}_n}}\Big[c\big(X_t,\gamma(X_{t})\big)\Big]\bigg|\to 0.
\end{align*}

For ease of notation we will first study the case where $t=2$.
\begin{align*}
&\sup_{\gamma\in \Gamma}\Big|E^{{\mathcal{T}}}\bigg[c\big(X_2,\gamma(X_{2})\big)\bigg]-E^{{\mathcal{T}_n}}\bigg[c\big(X_2,\gamma(X_{2})\big)\bigg]\Big|\\
&=\sup_{\gamma\in \Gamma}\big|\int \mathcal{T}(dx_1|x_0,\gamma(x_0))\mathcal{T}(dx_2|x_1,\gamma(x_1))c(x_,\gamma(x_{2}))\\
&\quad-\int \mathcal{T}_n(dx_1|x_0,\gamma(x_0))\mathcal{T}_n(dx_2|x_1,\gamma(x_{1})) c(x_2,\gamma(x_{2}))\big|.
\end{align*}
To show that above term goes to 0, we use a lemma parallel to Lemma A.2 in the paper.

\begin{lemma}
Suppose we have a uniformly bounded family of functions $\{f_n^\gamma : \mathds{X} \to \R, \gamma\in\Gamma_{eq}, n>0\}$ such that $\|f_n^\gamma\|_\infty<C$ for all $\gamma \in \Gamma_{eq}$ and for all $n>0$ for some $C<\infty$.

Further suppose we have another uniformly bounded family of functions $\{f^\gamma : \mathds{X} \to \R, \gamma\in\Gamma_{eq}\}$ such that $\|f^\gamma\|_\infty<C$ for all $\gamma \in \Gamma_{eq}$ for some $C<\infty$. Under the following assumptions,
\begin{itemize}
\item[(i)] 
For any $x_n \to x$
\begin{align}
&\sup_{\gamma \in \Gamma_{eq}}\big|f_n^\gamma(x_n)-f^\gamma(x)\big|\to 0\\
&\sup_{\gamma \in \Gamma_{eq}}\big|f^\gamma(x_n)-f^\gamma(x)\big|\to 0. \label{f2_fully}
\end{align}

\item[(ii)]
 $\mathcal{T}_n(\cdot|x_n,u_n)$ converges weakly to $\mathcal{T}(\cdot|x,u)$ for any $(x_n,u_n)\to (x,u)$.
\item[(iii)]
$\mathcal{T}(\cdot|x,u)$ is weakly continuous in $(x,u)$.
\item[(iv)] 
$\mathds{U}$ is compact.
\end{itemize}

 Then for some $x_t^n \to x_t$
\begin{align}\label{f3_fully}
&\sup_{\gamma \in \Gamma_{eq}} \bigg|\int \mathcal{T}_n(dx_{t+1}|x_t^n,\gamma(x_t^n))f_n^\gamma(x_{t+1}) - \int \mathcal{T}(dx_{t+1}|x_t,\gamma(x_t))f^\gamma(x_{t+1})\bigg| \to 0.
\end{align}
\end{lemma}
\proof
Using the same steps as in Lemma A.2 we can show that 
for any given compact set $K \subset \mathds{X}$, and $\epsilon>0$ there is a finite set of continuous functions $\mathds{F}:=\{f^1,\dots,f^N\}$ so that for any $\gamma$, there is $f^i \in \mathds{F}$ with 
\[\sup_{x \in K}|f^\gamma(x)-f^i(x) |\leq \epsilon.\] 

For the same $\epsilon>0$, the same $f^i \in \mathds{F}$ and the chosen compact set $K$, we can also make $\sup_{x\in K} |f_n^\gamma(x)-f^i(x)|\leq 3\epsilon/2$ for large enough $n$ and for any $\gamma \in \Gamma_{eq}$.

We also that the set of measures $S:=\cup_{\gamma \in \Gamma_{eq}} S_\gamma$ is weakly compact 
where 
\begin{align*}
S_\gamma=\{\mathcal{T}_n(\cdot|x_t^n,\gamma(x_t^n)): \mathcal{T}_n(\cdot|x_t^n,\gamma(x_t^n))\to \mathcal{T}(\cdot|x_t,\gamma(x_t))\}.
\end{align*}

 Hence, for any $\epsilon >0$, there exists a compact set $K_\epsilon$ such that for all $n$ and uniformly for all $\gamma \in \Gamma$, 
\[\int_{K_\epsilon} \mathcal{T}_n(dx_1|x_t^n,\gamma(x_t^n)) \geq 1-\epsilon.\]

Using these, we again follow the same steps as in the proof of Lemma A.2:
\begin{align*}
&\sup_{\gamma \in \Gamma_{eq}} \bigg|\int \mathcal{T}_n(dx_{t+1}|x_t^n,\gamma(x_t^n))f_n^\gamma(x_{t+1}) - \int \mathcal{T}(dx_{t+1}|x_t,\gamma(x_t))f^\gamma(x_{t+1})\bigg|\\
&\leq\sup_{\gamma \in \Gamma_{eq}}\big|\int_{\mathds{X}\setminus K_\epsilon} \mathcal{T}_n(dx_{t+1}|x_t^n,\gamma(x_t^n))f_n^\gamma(x_{t+1})- \int_{\mathds{X}\setminus K_\epsilon} \mathcal{T}(dx_{t+1}|x_t,\gamma(x_t))f^\gamma(x_{t+1})\big|\\
&\qquad+\sup_{\gamma \in \Gamma_{eq}}\big|\int_{ K_\epsilon} \mathcal{T}_n(dx_{t+1}|x_t^n,\gamma(x_t^n))f_n^\gamma(x_{t+1})- \int_{ K_\epsilon} \mathcal{T}(dx_{t+1}|x_t,\gamma(x_t))f^\gamma(x_{t+1})\big|\\
&\leq 2\epsilon C + \sup_{\gamma \in \Gamma_{eq}}\big|\int_{K_\epsilon} \mathcal{T}_n(dx_{t+1}|x_{t}^n,\gamma(x_t^n))\big(f_n^\gamma(x_{t+1}) -f^i(x_{t+1})\big)\\
&\qquad\qquad\qquad + \int_{K_\epsilon} \mathcal{T}_n(dx_{t+1}|x_{t}^n,\gamma(x_t^n))f^i(x_{t+1}) - \int_{K_\epsilon} \mathcal{T}(dx_{t+1}|x_t,\gamma(x_t))f^i(x_{t+1})\\
&\qquad\qquad\qquad\qquad + \int_{K_\epsilon} \mathcal{T}(dx_{t+1}|x_t,\gamma(x_t))\big(f^i(x_{t+1})-f^\gamma(x_{t+1})\big)\big|\\
&\leq 2\epsilon C + \sup_{\gamma \in \Gamma_{eq}} \big| \int_{K_\epsilon} \mathcal{T}_n(dx_{t+1}|x_{t}^n,\gamma(x_t^n))f^i(x_{t+1})\\
&\qquad\qquad\qquad\qquad - \int_{K_\epsilon} \mathcal{T}(dx_{t+1}|x_t,\gamma(x_t))f^i(x_{t+1})\big| + 5\epsilon/2 \leq 4\epsilon C + 7\epsilon/2
\end{align*}
where $C$ is the uniform bound of $f_n^\gamma$ and $f^i(x_{t+1})$ is chosen according to the discussion above such that $f^i$ is $\epsilon$ close to $f^\gamma(x_{t+1})$ and the same $f^i$ is $3\epsilon/2$ close to $f_n^\gamma(x_{t+1})$. 

At the last step, we used the fact that $\mathcal{T}_n(dx_{t+1}|x_{t}^n,\gamma(x_t^n))$ converges weakly to $\mathcal{T}(dx_{t+1}|x_{t},\gamma(x_t))$ uniformly over $\Gamma_{eq}$ as $\Gamma_{eq}$ is equicontinuous.
 As $\epsilon$ is arbitrary, the result follows.
\endproof


Now we go back to
\begin{align*}
&\sup_{\gamma\in \Gamma}\Big|E^{{\mathcal{T}}}\bigg[c\big(X_2,\gamma(X_{2})\big)\bigg]-E^{{\mathcal{T}_n}}\bigg[c\big(X_2,\gamma(X_{2})\big)\bigg]\Big|\\
&=\sup_{\gamma\in \Gamma}\big|\int \mathcal{T}(dx_1|x_0,\gamma(x_0))\mathcal{T}(dx_2|x_1,\gamma(x_1))c(x_2,\gamma(x_{2}))\\
&\quad-\int \mathcal{T}_n(dx_1|x_0,\gamma(x_0))\mathcal{T}_n(dx_2|x_1,\gamma(x_{1}))c(x_2,\gamma(x_{2}))\big|.
\end{align*}
The previous lemma can be used to show that this term converges to $0$.

\subsection{Proof for (\ref{k_conv}) }\label{tech_operator_conv}
We focus on the discounted optimality equation for $\mathcal{T}_n$ for some initial point $x_0^n$ where $x_0^n\to x_0$:
\begin{align}\label{tn_dcoe}
J_\beta^*(\mathcal{T}_n,x_0^n)=c(x_0^n,\gamma_n^*(x_0^n))+\beta \int J_\beta^*(\mathcal{T}_n,x_1)\mathcal{T}_n(dx_1|x_0^n,\gamma_n^*(x_0^n)).
\end{align}
Since the action space $\mathds{U}$ is compact, we can find a subsequence $n_k$ such that $\gamma^*_{n_k}(x_0^{n_k})\to u^*$ for some $u^*\in\mathds{U}$. Taking the limit $k\to \infty$ in (\ref{tn_dcoe}) and using  Theorem \ref{fully_weak_suff_cont} (continuity) we get

\begin{align}\label{tn_dcoe2}
J_\beta^*(\mathcal{T},x_0)=c(x_0,u^*)+\beta \int J_\beta^*(\mathcal{T},x_1)\mathcal{T}(dx_1|x_0,u^*).
\end{align}

Hence $u^*$ satisfies DCOE for $\mathcal{T}$ and is an optimal action for $x_0$. In particular, any convergent subsequence of $\gamma^*_n(x_0^n)$ converges to an optimal action for $x_0$. With this observation, we claim that $T_n^k(J_\beta^*(\mathcal{T}))(x_n)\to J_\beta^*(\mathcal{T},x))$ for any fixed $k<\infty$ as $n\to\infty$ for some $x_n\to x$ where $T_n^k$ denotes the operator $T_n$ is applied $k$ consecutive times. To show this, we follow an inductive approach. For $k=1$, we write
\begin{align*}
T_n(J_\beta^*(\mathcal{T}))(x_n)=c(x_n,\gamma^*_n(x_n))+ \beta\int J_\beta^*(\mathcal{T},x_1)\mathcal{T}(dx_1|x_n,\gamma^*_n(x_n)).
\end{align*}
Suppose $\lim T_n(J_\beta^*(\mathcal{T}))(x_n)\neq J_\beta^*(\mathcal{T},x)$ so that there exists a subsequence $n_m$ and an $\epsilon>0$ for which $| T_{n_m}(J_\beta^*(\mathcal{T}))(x_{n_m})- J_\beta^*(\mathcal{T},x)|>\epsilon$ for all m. Since $\mathds{U}$ is compact, there exists a further subsequence $n_{m'}$ such that $\gamma^*_{n_{m'}}(x_{n_{m'}})\to u$ for some $u\in\mathds{U}$ and as we observed before $u$ is an optimal action for $x$ under the kernel $\mathcal{T}$. Hence 
\begin{align*}
&\lim_{m'\to\infty}T_{n_{m'}}(J_\beta^*(\mathcal{T}))(x_{n_{m'}})\\
&=\lim_{m'\to\infty}c(x_{n_{m'}},\gamma^*_{n_{m'}}(x_{n_{m'}}))+ \beta\int J_\beta^*(\mathcal{T},x_1)\mathcal{T}(dx_1|x_{n_{m'}},\gamma^*_{n_{m'}}(x_{n_{m'}})\\
&=c(x,u)+ \beta\int J_\beta^*(\mathcal{T},x_1)\mathcal{T}(dx_1|x,u)=J_\beta^*(\mathcal{T},x)
\end{align*}
where the last step follows from the observation that $u$ is optimal for $x$ under $\mathcal{T}$. Thus, we reach a contradiction and can conclude that $T_n(J_\beta^*(\mathcal{T}))(x_n)\to J_\beta^*(\mathcal{T},x)$. Now assume that it also holds for $k-1$ so that $T_n^{k-1}(J_\beta^*(\mathcal{T}))(x_n)\to J_\beta^*(\mathcal{T},x)$. We write 
\begin{align*}
T_n^k(J_\beta^*(\mathcal{T}))(x_n)=c(x_n,\gamma^*_n(x_n))+ \beta\int T_n^{k-1}(J_\beta^*(\mathcal{T}))(x_1)\mathcal{T}(dx_1|x_n,\gamma^*_n(x_n)).
\end{align*}
Following a similar contradiction argument with the fact that $T_n^{k-1}(J_\beta^*(\mathcal{T}))(x_n)\to J_\beta^*(\mathcal{T},x)$ and using  \cite[Theorem 3.5]{Langen81} or \cite[Theorem 3.5]{serfozo82} (weak convergence with varying functions), we can conclude that 
\begin{align}\label{robust_oper}
\lim_{n\to\infty}T_n^k(J_\beta^*(\mathcal{T}))(x_n)= J_\beta^*(\mathcal{T},x),\quad \forall k<\infty.
\end{align}

\subsection{Proof for (\ref{robust_oper_set})} \label{tech_operator_conv_set}
We give a proof sketch building on Section \ref{tech_operator_conv}. Define the discounted optimality equation for $\mathcal{T}_n$ for some initial point $x_0$.
\begin{align}\label{tn_dcoe_set}
J_\beta^*(\mathcal{T}_n,x_0)=c(x_0,\gamma_n^*(x_0))+\beta \int J_\beta^*(\mathcal{T}_n,x_1)\mathcal{T}_n(dx_1|x_0,\gamma_n^*(x_0)).
\end{align}
Since the action space $\mathds{U}$ is compact, we can find a subsequence $n_k$ such that $\gamma^*_{n_k}(x_0)\to u^*$ for some $u^*\in\mathds{U}$. Taking the limit $k\to \infty$ in (\ref{tn_dcoe_set}) and using  Theorem \ref{fully_set_suff_cont} (continuity) we get

\begin{align}\label{tn_dcoe2_set}
J_\beta^*(\mathcal{T},x_0)=c(x_0,u^*)+\beta \int J_\beta^*(\mathcal{T},x_1)\mathcal{T}(dx_1|x_0,u^*).
\end{align}

Hence $u^*$ satisfies DCOE for $\mathcal{T}$ and is an optimal action for $x_0$. In particular, any convergent subsequence of $\gamma^*_n(x_0)$ converges to an optimal action for $x_0$. With this observation, we claim that $T_n^k(J_\beta^*(\mathcal{T}))(x)\to J_\beta^*(\mathcal{T},x))$ for any fixed $k<\infty$ as $n\to\infty$ for any $x$ where $T_n^k$ denotes the operator $T_n$ applied $k$ consecutive times. This can be shown by the same technique we use in Section \ref{tech_operator_conv} using \cite[Theorem 20]{royden} (setwise convergence with varying functions).

\bibliographystyle{plain}

\bibliography{references_acc,SerdarBibliography_acc,SerdarBibliography}

\begin{thebibliography}{10}

\bibitem{arruda2012}
A.~Almudevar. and E.~F. Arruda.
\newblock Optimal approximation schedules for a class of iterative algorithms,
  with an application to multigrid value iteration.
\newblock {\em IEEE Transactions on Automatic Control}, 57:3132--3146, 2012.

\bibitem{arruda2013}
E.~F. Arruda, F.~Ourique, J.~Lacombe, and A.~Almudevar.
\newblock Accelerating the convergence of value iteration by using partial
  transition functions.
\newblock {\em European Journal of Operational Research}, 229:190--198, 2013.

\bibitem{basbern}
T.~Ba\c{s}ar and P.~Bernhard.
\newblock {\em H-infinity Optimal Control and Related Minimax Design Problems:
  A Dynamic Game Approach}.
\newblock Birkh\"auser, Boston, MA, 1995.

\bibitem{benavoli2011robust}
A.~Benavoli and L.~Chisci.
\newblock Robust stochastic control based on imprecise probabilities.
\newblock {\em IFAC Proceedings Volumes}, 44(1):4606--4613, 2011.

\bibitem{bertsekas78}
D.~P. Bertsekas and S.~Shreve.
\newblock {\em Stochastic Optimal Control: The Discrete Time Case}.
\newblock Academic Press, New York, 1978.

\bibitem{billingsley1961statistical}
P.~Billingsley.
\newblock Statistical methods in {M}arkov chains.
\newblock {\em The Annals of Mathematical Statistics}, pages 12--40, 1961.

\bibitem{Billingsley}
P.~Billingsley.
\newblock {\em Convergence of Probability Measures}.
\newblock Wiley, New York, 1968.

\bibitem{BillingsleyProbMeasure}
P.~Billingsley.
\newblock {\em Probability and Measure}.
\newblock Wiley, 3rd ed.), New York, 1995.

\bibitem{blanchet2016}
J.~Blanchet and K.~Murthy.
\newblock Quantifying distributional model risk via optimal transport.
\newblock {\em SSRN Electronic Journal}, 04 2016.

\bibitem{boel2002robustness}
R.~K. Boel, M.~R. James, and I.~R. Petersen.
\newblock Robustness and risk-sensitive filtering.
\newblock {\em IEEE Transactions on Automatic Control}, 47(3):451--461, 2002.

\bibitem{BorkarRealization}
V.~S. Borkar.
\newblock White-noise representations in stochastic realization theory.
\newblock {\em SIAM J. on Control and Optimization}, 31:1093--1102, 1993.

\bibitem{Dean18}
S.~Dean, H.~Mania, N.~Matni, B.~Recht, and S.~Tu.
\newblock On the sample complexity of the linear quadratic regulator.
\newblock {\em arXiv preprint arXiv:1710.01688v2}, 2018.

\bibitem{Devroye85}
L.~Devroye and L.~Gy\"orfi.
\newblock {\em Non-parametric Density Estimation: The $L_1$ View}.
\newblock John Wiley, New York, 1985.

\bibitem{Dudley02}
R.~M. Dudley.
\newblock {\em Real Analysis and Probability}.
\newblock Cambridge University Press, Cambridge, 2nd edition, 2002.

\bibitem{dupuis2000robust}
P.~Dupuis, M.~R. James, and I.~Petersen.
\newblock Robust properties of risk-sensitive control.
\newblock {\em Mathematics of Control, Signals and Systems}, 13(4):318--332,
  2000.

\bibitem{FeKaZg_19}
P.~Kasyanov E.~Feinberg and M.~Zgurovsky.
\newblock On continuity of transition probabilities in belief mdps with general
  state and action spaces.
\newblock {\em arXiv preprint arXiv:1903.11629}, 2019.

\bibitem{erdogan2005}
E.~Erdo\u{g}an and G.~N. Iyengar.
\newblock Ambiguous chance constrained problems and robust optimization.
\newblock {\em Mathematical Programming}, 107(1-2):37--61, 2005.

\bibitem{esfahani2015}
P.~M. Esfahani and D.~Kuhn.
\newblock Data-driven distributionally robust optimization using the
  {W}asserstein metric: Performance guarantees and tractable reformulations.
\newblock {\em Mathematical Programming}, pages 1--52, 2017.

\bibitem{FeKaZg14}
Eugene~A. Feinberg, Pavlo~O. Kasyanov, and Michael~Z. Zgurovsky.
\newblock Partially observable total-cost markov decision processes with weakly
  continuous transition probabilities.
\newblock {\em Mathematics of Operations Research}, 41(2):656--681, 2016.

\bibitem{Ghosh}
J.~K. Ghosh and R.~V. Ramamoorthi.
\newblock {\em Bayesian Nonparametrics}.
\newblock Springer, New York, 2003.

\bibitem{gihman2012controlled}
I.~I. Gihman and A.~V. Skorohod.
\newblock {\em Controlled stochastic processes}.
\newblock Springer Science \& Business Media, 2012.

\bibitem{gossner2008entropy}
O.~Gossner and T.~Tomala.
\newblock Entropy bounds on {B}ayesian learning.
\newblock {\em Journal of Mathematical Economics}, 44(1):24--32, 2008.

\bibitem{GrayInfo}
R.~M. Gray.
\newblock {\em Entropy and Information Theory}.
\newblock Springer-Verlag, New York, 1990.

\bibitem{GyKo07}
L.~Gy\"orfi and M.~Kohler.
\newblock Nonparametric estimation of conditional distributions.
\newblock {\em IEEE Transactions Information Theory}, 53:1872--1879, May 2007.

\bibitem{hajek}
B.~Hajek.
\newblock {\em Random Processes for Engineers}.
\newblock Cambridge University Press, 2015.

\bibitem{hansen2001robust}
L.~P. Hansen and T.~J. Sargent.
\newblock Robust control and model uncertainty.
\newblock {\em American Economic Review}, 91(2):60--66, 2001.

\bibitem{HernandezLermaMCP}
O.~Hernandez-Lerma and J.~Lasserre.
\newblock {\em Discrete-time {M}arkov control processes}.
\newblock Springer, 1996.

\bibitem{hernandezlasserre1999further}
O.~Hern{\'a}ndez-Lerma and J.~B. Lasserre.
\newblock {\em Further topics on discrete-time {M}arkov control processes}.
\newblock Springer, 1999.

\bibitem{Iyengar2005}
G.~N. Iyengar.
\newblock Robust dynamic programming.
\newblock {\em Mathematics of Operations Research}, 30(2):257--280, 2005.

\bibitem{jacobson1973optimal}
D.~Jacobson.
\newblock Optimal stochastic linear systems with exponential performance
  criteria and their relation to deterministic differential games.
\newblock {\em IEEE Transactions on Automatic control}, 18(2):124--131, 1973.

\bibitem{KSY2019scl}
A.~D. Kara, N.~Saldi, and S.~Y{\"u}ksel.
\newblock Weak {F}eller property of non-linear filters.
\newblock {\em Systems \& Control Letters}, 134:104512, 2019.

\bibitem{Prior2017}
A.~D. Kara and S.~Y\"uksel.
\newblock Robustness to incorrect priors in partially observed stochastic
  control.
\newblock {\em SIAM Journal on Control and Optimization}, 57(3):1929--1964,
  2019.

\bibitem{Kleptsyna2016}
M.~L. Kleptsyna and A.~Yu. Veretennikov.
\newblock On robustness of discrete time optimal filters.
\newblock {\em Mathematical Methods of Statistics}, 25(3):207--218, 2016.

\bibitem{lam2016}
H.~Lam.
\newblock Robust sensitivity analysis for stochastic systems.
\newblock {\em Mathematics of Operations Research}, 41(4):1248--1275, 2016.

\bibitem{Langen81}
H.J. Langen.
\newblock Convergence of dynamic programming models.
\newblock {\em Mathematics of Operations Research}, 6(4):493--512, 1981.

\bibitem{muller1997does}
A.~M{\"u}ller.
\newblock How does the value function of a markov decision process depend on
  the transition probabilities?
\newblock {\em Mathematics of Operations Research}, 22(4):872--885, 1997.

\bibitem{Ghaoui2005}
A.~Nilim and L.~El Ghaoui.
\newblock Robust control of {M}arkov decision processes with uncertain
  transition matrices.
\newblock {\em Operations Research}, 53(5):780--798, 2005.

\bibitem{oksendal2014forward}
B.~{\O}ksendal and A.~Sulem.
\newblock Forward--backward stochastic differential games and stochastic
  control under model uncertainty.
\newblock {\em Journal of Optimization Theory and Applications}, 161(1):22--55,
  2014.

\bibitem{Par67}
K.R. Parthasarathy.
\newblock {\em Probability Measures on Metric Spaces}.
\newblock AMS Bookstore, 1967.

\bibitem{dupuis2000kernel}
I.~Petersen, M.~R. James, and P.~Dupuis.
\newblock Minimax optimal control of stochastic uncertain systems with relative
  entropy constraints.
\newblock {\em IEEE Transactions on Automatic Control}, 45(3):398--412, 2000.

\bibitem{dai1996connections}
P.~Dai Pra, L.~Meneghini, and W.~J. Runggaldier.
\newblock Connections between stochastic control and dynamic games.
\newblock {\em Mathematics of Control, Signals and Systems}, 9(4):303--326,
  1996.

\bibitem{royden}
H.~L. Royden.
\newblock {\em Real Analysis}.
\newblock Macmillan, New York, 1968.

\bibitem{saldi2016markov}
N.~Saldi, T.~Ba{\c{s}}ar, and M.~Raginsky.
\newblock Markov-{N}ash equilibria in mean-field games with discounted cost.
\newblock {\em arXiv preprint arXiv:1612.07878}, 2016.

\bibitem{SLYbook}
N.~Saldi, T.~Linder, and S.~Yu\"ksel.
\newblock {\em Finite Approximations in Discrete-time Stochastic Control:
  Quantized Models and Asymptotic Optimality.}
\newblock Birkhäuser, 2018.

\bibitem{SYL2016near}
N.~Saldi, S.~Y{\"u}ksel, and T.~Linder.
\newblock Near optimality of quantized policies in stochastic control under
  weak continuity conditions.
\newblock {\em Journal of Mathematical Analysis and Applications},
  435(1):321--337, 2016.

\bibitem{SaYuLi17}
N.~Saldi, S.~Y\"uksel, and T.~Linder.
\newblock Asymptotic optimality of finite approximations to {M}arkov decision
  processes with {B}orel spaces.
\newblock {\em Math. Oper. Res.}, pages 1--34, March 2017.

\bibitem{savkin1996robust}
A.~V. Savkin and I.~R. Petersen.
\newblock Robust control of uncertain systems with structured uncertainty.
\newblock {\em Journal of Mathematical Systems, Estimation, and Control},
  6(3):1--14, 1996.

\bibitem{Schal}
M.~Sch\"al.
\newblock Conditions for optimality in dynamic programming and for the limit of
  n-stage optimal policies to be optimal.
\newblock {\em Z. Wahrscheinlichkeitsth}, 32:179--296, 1975.

\bibitem{serfozo82}
R.~Serfozo.
\newblock Convergence of lebesgue integrals with varying measures.
\newblock {\em Sankhy{\=a}: The Indian Journal of Statistics, Series A}, pages
  380--402, 1982.

\bibitem{sun2015}
H.~Sun and H.~Xu.
\newblock Convergence analysis for distributionally robust optimization and
  equilibrium problems.
\newblock {\em Mathematics of Operations Research}, 41:377--401, 07 2015.

\bibitem{tzortzis2015dynamic}
I.~Tzortzis, C.D. Charalambous, and T.~Charalambous.
\newblock Dynamic programming subject to total variation distance ambiguity.
\newblock {\em SIAM Journal on Control and Optimization}, 53(4):2040--2075,
  2015.

\bibitem{ugrinovskii1998robust}
V.~A. Ugrinovskii.
\newblock Robust {H}-infinity control in the presence of stochastic
  uncertainty.
\newblock {\em International Journal of Control}, 71(2):219--237, 1998.

\bibitem{villani2008optimal}
C.~Villani.
\newblock {\em Optimal transport: old and new}.
\newblock Springer, 2008.

\bibitem{wheeden77}
R.~L. Wheeden and A.~Zygmund.
\newblock {\em Measure and Integral}.
\newblock Marcel Dekker, New York, 1977.

\bibitem{Rustem2012}
W.~Wiesemann, D.~Kuhn, and B.~Rustem.
\newblock Robust {M}arkov decision processes.
\newblock {\em Mathematics of Operations Research}, 38(1):153--183, 2012.

\bibitem{xu_mannor}
H.~Xu and S.~Mannor.
\newblock Distributionally robust {M}arkov decision processes.
\newblock In J.~D. Lafferty, C.~K.~I. Williams, J.~Shawe-Taylor, R.~S. Zemel,
  and A.~Culotta, editors, {\em Advances in Neural Information Processing
  Systems 23}, pages 2505--2513. Curran Associates, Inc., 2010.

\bibitem{yuksel12:siam}
S.~Y\"uksel and T.~Linder.
\newblock Optimization and convergence of observation channels in stochastic
  control.
\newblock {\em SIAM J. on Control and Optimization}, 50:864--887, 2012.

\bibitem{zhou1996robust}
K.~Zhou, J.~C. Doyle, and K.~Glover.
\newblock {\em Robust and optimal control}, volume~40.
\newblock Prentice-Hall, 1996.

\end{thebibliography}

\end{document}